\documentclass{TIA}

\begin{document}

\title{A Cyber-HIL for Investigating Control Systems\\ in Ship Cyber Physical Systems \\under Communication Issues and Cyber Attacks}
\author{
	Linh Vu,~\IEEEmembership{Student Member,~IEEE,}
	Lam Nguyen,~\IEEEmembership{Member,~IEEE,}
	Mahmoud Abdelrahman,~\IEEEmembership{Student Member,~IEEE,}
	Tuyen Vu,~\IEEEmembership{Member,~IEEE,} Osama Mohammed,~\IEEEmembership{Fellow Member,~IEEE}
}


\maketitle

\begin{abstract}
	This paper presents a novel Cyber-Hardware-in-the-Loop (Cyber-HIL) platform for assessing control operation in ship cyber-physical systems. The proposed platform employs cutting-edge technologies, including Docker containers, real-time simulator $OPAL-RT$, and network emulator $ns3$, to create a secure and controlled testing and deployment environment for investigating the potential impact of cyber attack threats on ship control systems. Real-time experiments were conducted using an advanced load-shedding controller as a control object in both synchronous and asynchronous manners, showcasing the platform's versatility and effectiveness in identifying vulnerabilities and improving overall Ship Cyber Physical System (SCPS) security. Furthermore, the performance of the load-shedding controller under cyber attacks was evaluated by conducting tests with man-in-the-middle (MITM) and denial-of-service (DoS) attacks. These attacks were implemented on the communication channels between the controller and the simulated ship system, emulating real-world scenarios. The proposed Cyber-HIL platform provides a comprehensive and effective approach to test and validate the security of ship control systems in the face of cyber threats.
\end{abstract}


\section{Introduction}


\IEEEPARstart{T}{he} integration of cyber-physical systems (CPS) into ships has become a crucial aspect of the maritime industry's digital transformation, with the aim of optimizing the performance and enhancing the resilience of ship power systems \cite{Progoulakis2023, jmse9121384}. The Ship Cyber Physical System (SCPS) is a prime example of such integration, relying heavily on efficient communication networks for successful operation. These modern ship systems incorporate a multitude of highly accurate sensors, sophisticated controllers, and swift-acting actuators, which facilitate the instantaneous monitoring and management of ship systems during different operational tasks, including propulsion, navigation, cargo handling, and power distribution \cite{GieringDyck+2021+1081+1095}.

To ensure safe and efficient ship operations, the network interconnecting these components must offer reliable and timely communication. Hence, it is imperative to conduct a thorough and comprehensive assessment of the network's capabilities under challenging and realistic scenarios to maximize power continuity \cite{markle2019naval}. This is because any failure or compromise of a communication infrastructure component during regular or combat operations could trigger a cascading failure, which may result in severe damage to the physical components of the system \cite{KARAHALIOS20181, 9329185}. In such a scenario, there is a risk of losing control over the vessel, which could pose a serious safety hazard for the crew and compromise the success of the ship's mission \cite{boyes2017code}.

Ship Cyber Physical Systems (SCPS) research is typically categorized into two main areas: the cyber and physical domains. The cyber aspect of SCPS encompasses communication networks, software, and data management systems, which operate in a discrete manner. In contrast, the physical domain includes the mechanical, electrical, and electronic components that constitute the ship power systems, which operate continuously \cite{YAACOUB2020103201}. The need to guarantee security and reliability when integrating the cyber and physical domains has led to the development of various testing methodologies such as Hardware-in-the-loop (HIL) simulation \cite{9494202, 9096292, 9184969, 9526562}. HIL simulation involves connecting a physical system or component to a virtual or simulated environment for testing and validating purposes. This approach provides a cost-effective and safe way to test control algorithms and hardware components in a controlled environment, especially in ship systems.

Another important testing methodology in the context of SCPS is co-simulation \cite{4906565, MIC-2020-4-2}, which allows the cyber and physical domains to be simulated simultaneously in a coupled manner. Co-simulation involves linking different simulation tools (such as OPAL-RT, RTDS, and Typhoon HIL), each can be considered as a black box to simulate a particular aspect of the SCPS, in order to capture the interactions between the cyber and physical domains. These simulations can be operated in either synchronous or asynchronous ways. In synchronous simulations \cite{5619971, 9325305}, all the components are communicated using a common time step, with each component updating its state variables at the same time. Asynchronous simulations \cite{9046860, 8412096}, on the other hand, do not require a common time step and can update the state variables of each component at different times. The choice between synchronous and asynchronous simulations depends on several factors, such as the complexity of the power system, simulation objectives, and available computational resources. For SCPS, which often involves geographically dispersed, distributed components that communicate through different mediums, asynchronous simulation can be particularly useful for modeling the interactions between these components to identify the communication delays and disruptions that may occur in such networks.

Apart from the inherent issues such as communication delay, the current SCPS is also exposed to a high degree of vulnerability to cyber attacks. Due to the intricate network involved in SCPS, there exist various objectives for cyber attacks. For instance, electronic chart display and information system (ECDIS) \cite{8433151}, voyage data recorder (VDR) \cite{jmse11020267}, emergency position indication radio beacon (EPIRB) \cite{Costin2023CybersecurityOC}, vessel performance monitoring system, and port management information system \cite{osti_10166239} are all potential targets. The cyber attacks in SCPS can take many various forms, making them difficult to identify, detect, and mitigate. Some of the common cyber threats that can be unleashed on SCPS include phishing attacks \cite{9444397, 6497928}, ransomware attacks \cite{10.1007/978-981-19-6414-5_6}, denial-of-service (DoS) attacks \cite{9275344, 8985406}, malware attacks \cite{greenberg-2019}, and man-in-the-middle attacks \cite{6493150, 10069874}. These attacks can lead to severe consequences, such as loss of data, system shutdown, financial losses, and compromise of sensitive information. As such, it is crucial to implement robust security measures to safeguard the SCPS from cyber attacks.

\rev{Research on SCPS is motivated by the importance of robust networks in ship systems, potential consequences of network failures, the need for HIL testing, co-simulation (synchronous/asynchronous modes) with communication delay analysis, and addressing cyber attack issues. Advancing these areas is crucial for developing secure and resilient SCPS for the maritime industry.} In this paper, we present a novel approach to investigate the potential impact of cyber-attacks on ship control systems. To this end, we developed a secure and controlled testing and deployment environment. Real-time experiments were conducted using an ALS controller as a control object in both synchronous and asynchronous modes. The results demonstrate the platform's versatility and effectiveness in identifying potential vulnerabilities and improving the overall security of SCPS. Furthermore, the platform's high scalability makes it a valuable tool for handling large volumes of data and processing tasks in a highly distributed environment of SPCS.

\noindent \textbf{The following are the paper's main contributions:}

\begin{itemize}
	\item \rev{A container-based Cyber-HIL platform is developed to evaluate the operation of ship cyber-physical systems. The containers act as a gateway, linking power system devices to communication nodes, as well as can be virtual sources to deploy real cyber attacks. The inherent flexibility of this setup empowers the platform to efficiently incorporate a diverse range of communication network challenges, cyber attacks, and various types of controllers on SCPS, ensuring its adaptability and versatility in assessing system behavior and security.}
	\item {\rev{An advanced load-shedding controller is provided as the control object and integrated into the Cyber-HIL platform through the implementation of the Modbus TCP protocol for data exchange.}}
	\item {\rev{The Cyber-HIL platform will be utilized to assess the performance of the control system under various scenarios, including man-in-the-middle (MITM) and denial-of-service (DoS) attacks, as well as communication issues. This comprehensive evaluation aims to demonstrate the platform's effectiveness in assessing the vulnerability of ship power systems to potential cyber-attacks.}}
\end{itemize}

The remainder of this paper is organized as follows: In Section \ref{methodology}, we present our methodology, which includes the introduction of our proposed Cy-HIL platform in subsection \ref{CyHIL setup} and the description of the ALS controller as a control object in subsection \ref{ALS}. Section \ref{experiment} is divided into three subsections to assess the controller's performance under communication delay (\ref{comdelay1}, \ref{comdelay2}) and cyber attacks (\ref{cyberattack}).

\section{Methodology}
\label{methodology}
\subsection{The Cyber Hardware-in-the-loop (Cy-HIL) Setup for Ship Cyber Physical Systems}
\label{CyHIL setup}
\noindent The ship controllers require information about the ship's system from remote meters. This information is used to calculate control commands for controllable loads. The ship's communication infrastructure is used to collect data and send control signals. Therefore, not only the controller design, but also the data transmission process is critical to the performance of the ship operation. The speed, reliability, and security of the data transfer are essential factors that must be considered to ensure optimal control system performance. Any delay or interruption in data transmission can lead to inaccuracies in the control outputs, which can affect the safety and efficiency of the ship's operations.

\TIAFigs{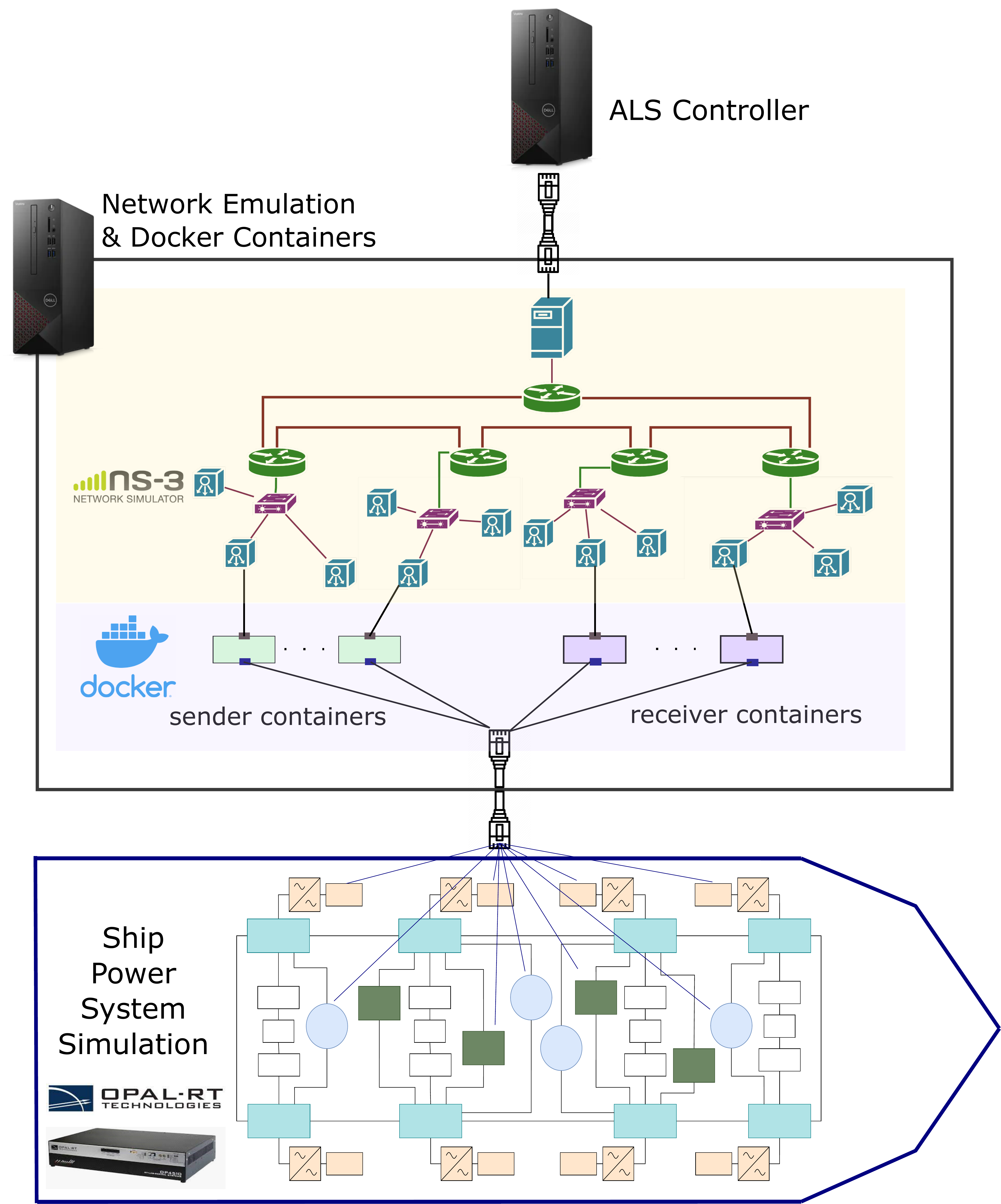}{Cyber Hardware-in-the-Loop setup.}{fig:cy_hil}{0.18}{}

In this study, a Cy-HIL setup for SCPS is presented to investigate the operation of the ALS controller under realistic and non-ideal cyber system conditions. The platform setup is shown in Fig.~\ref{fig:cy_hil} consisting of three major components: (1) $OPAL-RT$ simulates the real-time behavior of an MVAC four zones ship power system in \cite{esrdc1314}, (2) a Linux OS computer with $ns3$ simulation emulating the real-time behavior of the communication network and a Docker container cluster, and (3) a \rev{hardware} central controller. The containerized concept is used to establish scalable and flexible interfaces between $OPAL-RT$ and $ns3$.

\subsubsection{Real-time simulation of ship power systems in OPAL-RT}

\begin{figure*}[t]
	\centering
	\includegraphics[width=0.7\linewidth]{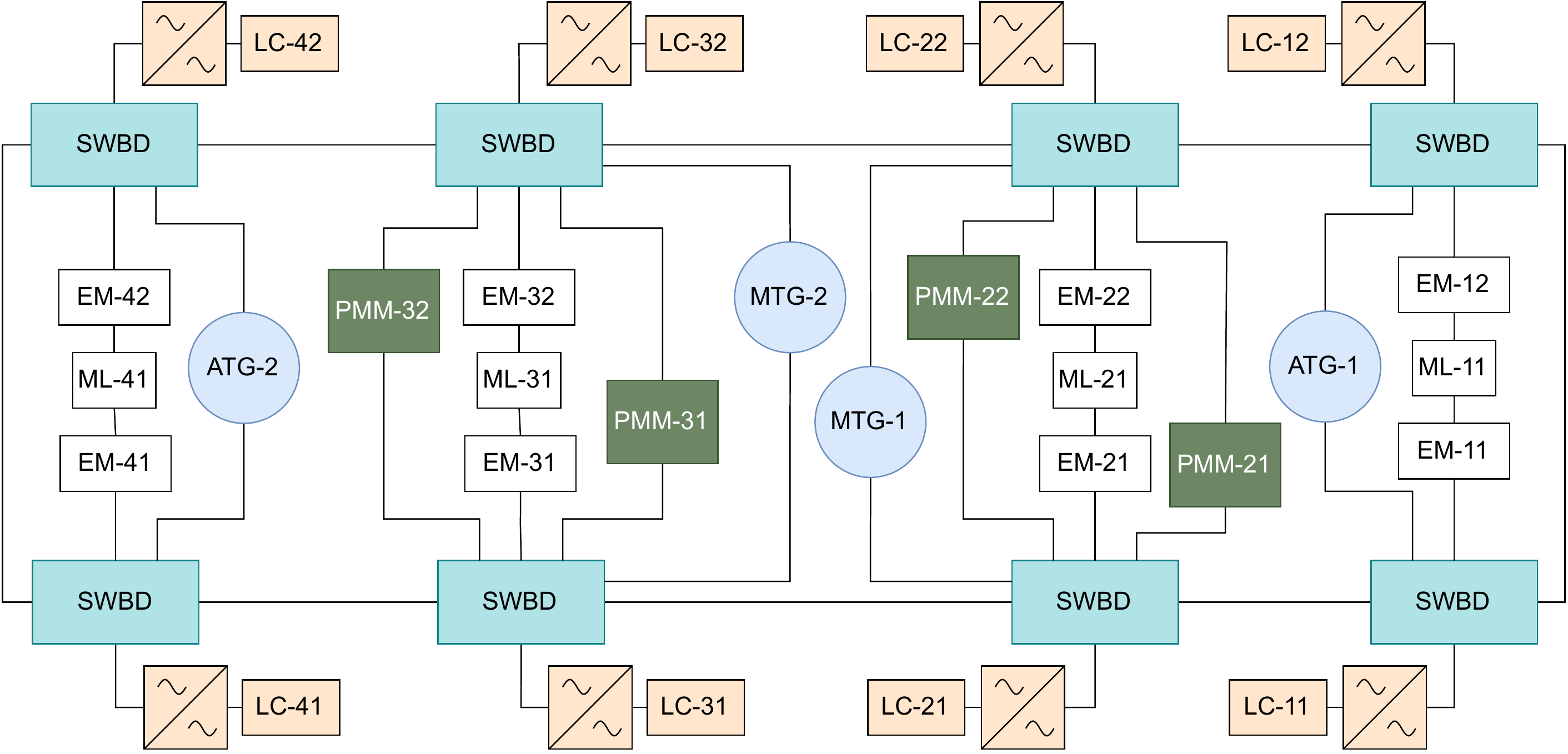}
	\caption{\rev{The diagram of a notional four-zone MVAC system.}}
	\label{fig:mvac}
\end{figure*}

\rev{The ship power system is modeled based upon the Medium \rev{Voltage Direct Alternating Current} (MVAC) notation for a four-zone ship system \cite{esrdc1314} as shown in Fig. \ref{fig:mvac} and is executed in real-time using the $OPAL-RT$ simulator. The MVAC system is divided into four zones, each comprising several modules as illustrated in Table \ref{tab:MVAC-component}, including four Propulsion Motor Modules (PMMs), eight Load Centers (LCs), four Mission Loads (MLs), two Main Turbine Generators (MTGs), two Auxiliary Turbine Generators (ATGs), eight Energy Magazines (EMs) and eight Switch Boards (SWBDs). Those components and their interconnections are modeled in Simulink and then loaded into $OPAL-RT$. This real-time simulator is connected to the communication network via a dedicated Ethernet port, enabling it to interface with external devices or control systems through a HIL setup to evaluate the performance of the entire complex cyber-physical systems. The exchange of internal signals with external systems occurs through the use of an asynchronous UDP/IP protocol. Although the ship power system is capable of operating independently without the supervision of a central controller, the optimal operation, stability, and reliability of the system cannot be guaranteed.}

\begin{table}[!htb]
	\caption{\rev{Component parameters of the MVAC ship system}}
	\label{tab:MVAC-component}
	\begin{tabular}{@{}llcc@{}}
		\toprule
		Symbol & Component                    & \multicolumn{1}{l}{Quantity} & \multicolumn{1}{l}{Power Rating (MW)} \\ \midrule
		PMM    & Propulsion motor modules     & 4                            & 15                                    \\
		LC     & Load centers                 & 8                            & 1                                     \\
		ML     & Mission loads                & 4                            & 1                                     \\
		MTG    & Main turbine generators      & 2                            & 35                                    \\
		ATG    & Auxiliary turbine generators & 2                            & 4.5                                   \\
		EM     & Energy magazines             & 8                            & 1                                     \\
		SWBD   & Switch boards                & 8                            & -                                     \\ \bottomrule
	\end{tabular}
\end{table}


\subsubsection{Network emulation in ns3}

The communication network within the simulated power system is emulated using $ns3$ \cite{Riley2010}, a discrete-event network simulator. With a modular implementation, $ns3$ features a core library that establishes the fundamental framework for the communication network, encompassing temporal objects, schedules, and events for the simulation. To achieve real-time communication emulation, $ns3$ runs on a dedicated server, utilizing system time for event scheduling.

\subsubsection{Containers as data buffers}

The Docker container \cite{merkel2014docker} is a lightweight and portable software package designed to encapsulate applications and their dependencies, ensuring consistent operation across diverse environments. It simplifies the deployment and management of complex software systems by providing an isolated environment. In this setup, Docker containers coexist with the $ns3$ network emulation system on a Linux host computer. As illustrated in Fig. \ref{fig:container}, the container acts as an intermediary between real-time simulation in $Opal-RT$ and $ns3$, equipped with a virtual network interface and access to shared memory on the host computer.

\begin{itemize}
	\item \rev{\textbf{The virtual network:} }The virtual network interface serves as a gateway for exchanging data among containers via the $ns3$ platform. Specifically, the virtual network interface is linked to Linux bridges that establish connectivity with the host operating system. In addition, tapping devices are attached to these bridges to intercept packets that traverse them and deliver them to user space where they can be accessed by $ns3$. A specialized $ns3$ NetDevice is connected to the network socket and transmits packets to a $ns3$ ghost node. The ghost node serves as a stand-in for the container in the $ns3$ simulation. Packets that enter through the network tap are transmitted through the corresponding net device, while packets that enter through the net device are transmitted through the network tap.
	\item \rev{\textbf{The Modbus TCP:} }The Modbus TCP protocol is used to transmit data between the central controller and the devices. Modbus TCP is frequently used in power systems for monitoring and controlling devices such as remote terminal units (RTUs), intelligent electronic devices (IEDs), and programmable logic controllers (PLCs). The protocol's ability to transmit data over Ethernet networks makes it well-suited for use in modern power system communication infrastructures. It is a master-slave protocol that enables a master device to communicate with multiple slave devices. Each container, which represents a device, is a Modbus TCP slave, while the central controller is a Modbus TCP master.
	\item \rev{\textbf{Shared memory:}} The Linux operating system offers a shared memory mechanism that allows multiple processes to concurrently access the same memory region, facilitating data sharing and interprocess communication and improving system efficiency and performance. Within the present setup, the combination of Docker containers and shared memory offers an advantageous approach to establishing a data buffer on shared memory. This data buffer serves the dual purposes of (i) collecting measurement signals from $Opal-RT$, transferring them to a container, and (ii) collecting control signals from the container and transferring them to $Opal-RT$.
\end{itemize}
\TIAFigs{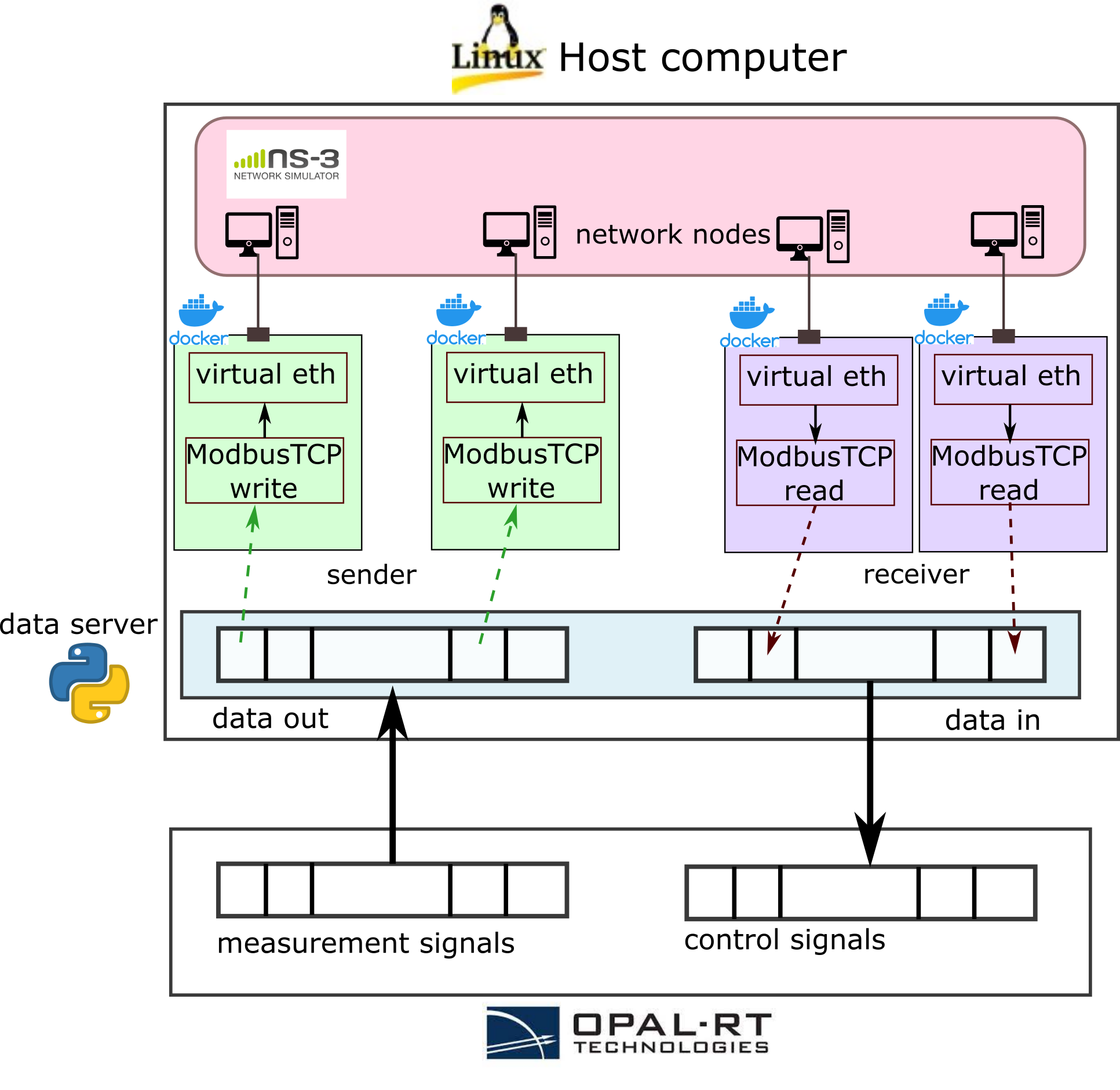}{Container structure.}{fig:container}{0.17}{}

The scalability of this design is its distinguishing feature. This is achieved through the use of a containerized approach that links each physical device in the system to a corresponding node in the communication network. In addition, the inclusion of shared memory processes emulates the actual functionality of sensor/local controller devices, facilitating the collection of data locally from physical systems and the use of communication infrastructures for remote data exchange and collection. This approach is not limited by communication hardware, resulting in increased system flexibility and adaptability.

\subsubsection{Central controller} The central controller is developed and deployed on a distinct computer, separate from the real-time simulator and the network emulator. The host computer, connected to the Linux computer via Ethernet, is integrated into the ship communication network as a node in $ns3$. Facilitated by a Modbus TCP master, the central controller can communicate with the devices simulated in $Opal-RT$ via network emulation. Operating independently and asynchronously from the remainder of the system, the controller operation enhances the realism of the test.

\rev{The Cy-HIL is provided at a system level and can be leveraged to assess the performance of a control system with one or multiple controllers. These controllers may either be hardware devices or virtually emulated within containers. By employing network emulation to separate the controllers, we can utilize existing software in the simulated nodes, minimizing the necessity for ad hoc simulation software and enhancing the accuracy of the results. Additionally, the design, programming, and implementation of the controllers become more manageable, flexible, and reusable.}

\subsection{Advanced Load-shedding for Ship Cyber Physical Systems}
\label{ALS}
\noindent The \rev{utilization} of a controller as a control object is \rev{crucial for assessing the communication performance of a prosed platform, especially when analyzing its behavior under communication delays or cyberattacks.} Therefore, this section presents an advanced load-shedding (ALS) control approach to evaluate the operational performance of the proposed Cy-HIL platform across different scenarios.

In electric ship power systems, loss of generators or line contingency can result in a shortage of power generation \cite{9917131, 9512317}. Since the ship power system has no \rev{interconnection} to neighboring utilities for power backup, shedding loads during power generation shortages becomes crucial to preserve critical loads and prevent widespread system outages. To enhance the resilience of ship power systems in such conditions, the ALS control algorithm is present to disconnect nonessential loads to maintain \rev{the} power balance between generation and load, thus reducing the risk of the ship system's collapse.

\TIAFigs{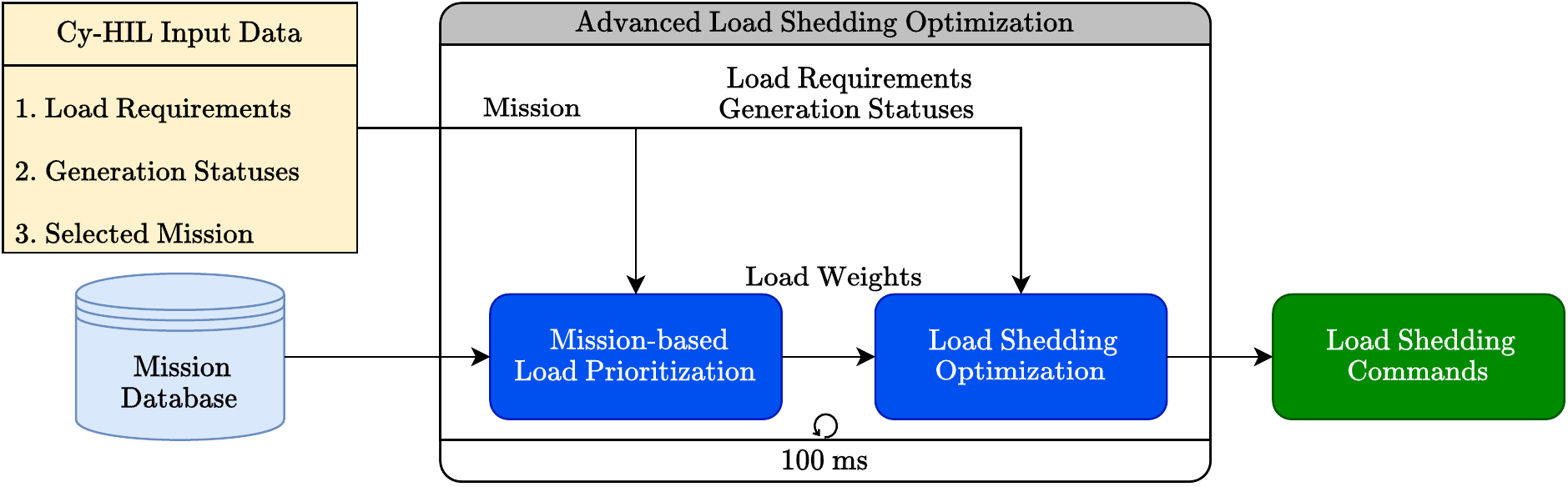}{Block diagram of mission-based load-shedding control.}{fig:mission_based}{0.455}{}

This ALS controller is mission-based and designed to shed loads based on the priority sets of loads in various missions during power shortage events. As illustrated in Fig. \ref{fig:mission_based}, the Cy-HIL platform \rev{periodically} updates the mission modes, system topology, the status of generators, and load requirements. The mission database contains load prioritization information for specific missions and the criticality of loads to the system. The ALS optimization algorithm calculates the load weights based on the operating mission and load prioritization, and then uses an optimization solver to generate load-shedding commands that align with the system's objectives. This process is repeated in real-time every $T_{s}$.

\rev
{
	As shown in (\ref{eq:Oi}), an operability metric denoted as  $O$ \cite{9512362, 5705696} is utilized to measure the resilience of load serving that contributes to mission effectiveness in a particular scenario over a defined period of time.
}

\rev{
\begin{equation}
	\begin{aligned}
		\label{eq:Oi}
		O         & = {\frac{{\int_{t_0}^{t_f}}{\sum_{i = 1}^{n_L}}{\hat{w}_i o_i^t} dt}{{\int_{t_0}^{t_f}}{\sum_{i = 1}^{n_L}}{\hat{w}_i o_i^{t*}} dt
		}}                                                                                                                                             \\
		\hat{w}_i & = w_i P_{i}^{L*} \quad \quad \quad \quad \quad \forall i \in n_L,
	\end{aligned}
\end{equation}
, where $n_L$ represents the number of loads, $t_0$ and $t_f$ indicate the start and end times of the event respectively. $P_i^{L*}$ denotes the rated power of load $i$, $w_i$ corresponds to the weight value of load $i$, $o_i^t$ signifies the actual operational status of load $i$ at time $t$, and $o_i^{t*}$ represents the required operational status of load $i$ at time $t$.
}

\rev{Building upon the concept of the aforementioned operability metric,} the load-shedding problem is approached by formulating it based on the load weights ($w_i$) to determine the loads and the amount of load to shed. The optimization problem's optimal variable is the operational status ($o_i$), which denotes the operability of each load $i$. For instance, if $o_i = 0$, the load $i$ is entirely shed, and if $o_i = 1$, load $i$ is served completely.

The objective function of the ALS problem is shown in (\ref{eq:Obj_central}), which includes two objectives:

\begin{itemize}
	\item The primary objective is to maximize the load operability metric \rev{at each given time step}.
	\item \rev{The secondary objective is to reduce the frequency of load switching by minimizing the total changes in the operational status between two consecutive time steps.}
\end{itemize}

\begin{maxi!}[3]
{o_i^L(t)}
{
{\frac{{\sum_{i = 1}^{n_L}}{\hat{w}_i o_i^L(t)}}{{\sum_{i = 1}^{n_L}}{\hat{w}_i o_i^{L*}}}} - \alpha \sum_{i = 1}^{n_L}{\lvert o_i^L(t) - o_i^L(t-1) \rvert}
}
{}{}
\label{eq:Obj_central}
\addConstraint{
\sum_{i=1}^{n_L}{P_i^{L*}(t)} o_i^L(t)}{ \leq (1-\beta)\sum_{g = 1}^{n_G}{P_g^G(t)}
}{}
\label{eq:st_balance}
\addConstraint{
r^L_{\text{min}} \leq \Delta P_i^L(t) \leq r^L_{\text{max}},}{}{\quad \forall i \in n_L
}
\label{eq:st_ramp_rate}
\addConstraint{
\Delta P_i^L(t) = P_i^{L*}(t) o_i^L(t) - P_i^{L*}(t-1) o_i^L(t-1),}{}{\quad \forall i \in n_L
}
\label{eq:st_ramp_define}
\addConstraint{
	0 \leq o_i^L(t) \leq 1,}{}{\quad \forall i \in n_L
}
\label{eq:st_operation_status}
\addConstraint{
	o_i^L(t) \in
	\begin{cases}
		\mathcal{Z} & \text{if load } i \text{ is the step-size load,} \\
		[0, 1]      & \text{if load } i \text{ is the continuous load}
	\end{cases}
}
\label{eq:st_operation_status_define}
\end{maxi!}
, where $\alpha$ is a constant coefficient to adjust the load switching weight compared to the primary objective, $\beta$ is a constant coefficient for backing up power generation capacity, $P_i^{L*}(t)$ and $o_t^L(t)$ are the reference power and operational status of load $i$ at time-step $t$, $P_i^{L*}(t-1)$ and $o_t^L(t-1)$ are the reference power and operational status of load $i$ at the previous time-step interval, $P_g^{G}(t)$ is the available power of generator $g$ at time-step $t$, $r_{min}^L$ and $r_{max}^L$ are the minimum and maximum shedding ramp-rate limitations, \rev{$n_G$ and $n_L$ are the number of generators and loads in the system, respectively.}

The objective function (\ref{eq:Obj_central}) is subject to several safety constraints (\ref{eq:st_balance} - \ref{eq:st_operation_status_define}) in order to ensure feasible solutions for the physical ship systems. Constraint (\ref{eq:st_balance}) imposes a limit on the total power demand from loads at each given time step, guaranteeing that it does not exceed the available generation capacity while also accounting for the reserved power of generators \rev{using the $\beta$ coefficient}. Constraints (\ref{eq:st_ramp_rate} - \ref{eq:st_ramp_define}) set limits on the amount of served load to be changed at each time-step to avoid sudden drops or rises in frequency and voltage of the ship systems. The operational status of loads is restricted by constraint (\ref{eq:st_operation_status}). The final constraint (\ref{eq:st_operation_status_define}) specifies that certain types of loads can only be shed in discrete increments, where $\mathcal{Z} = \{0:\Delta o_i:1\}$ and $\Delta o_i = 1/n$, with $n$ being the maximum number of commanded steps.

\section{Experimental Results}
\label{experiment}
\noindent \rev{With the Cy-HIL setup in Section \ref{CyHIL setup}, a comprehensive range of tests were conducted on the communication network to evaluate the performance of SCPS. These tests encompassed various scenarios, including examinations of network issues such as changes in data transmission rates and network congestion, as well as emulations of potential cyber attacks, such as DoS and MITM attacks. The testing involved real-time simulations of the communication network and SCPS, achieved through $ns3$ and $Opal-RT$. However, it's worth noting that the controller was implemented on a separate computer (Intel(R) Core(TM) i7-10700 CPU) and operated independently from the rest of the system. To ensure real-time execution of the load-shedding controller, Simulink Desktop Real-Time was employed for its implementation in Simulink.}
\subsection{Communication Network Performance}
\label{comdelay1}

\TIAFigs{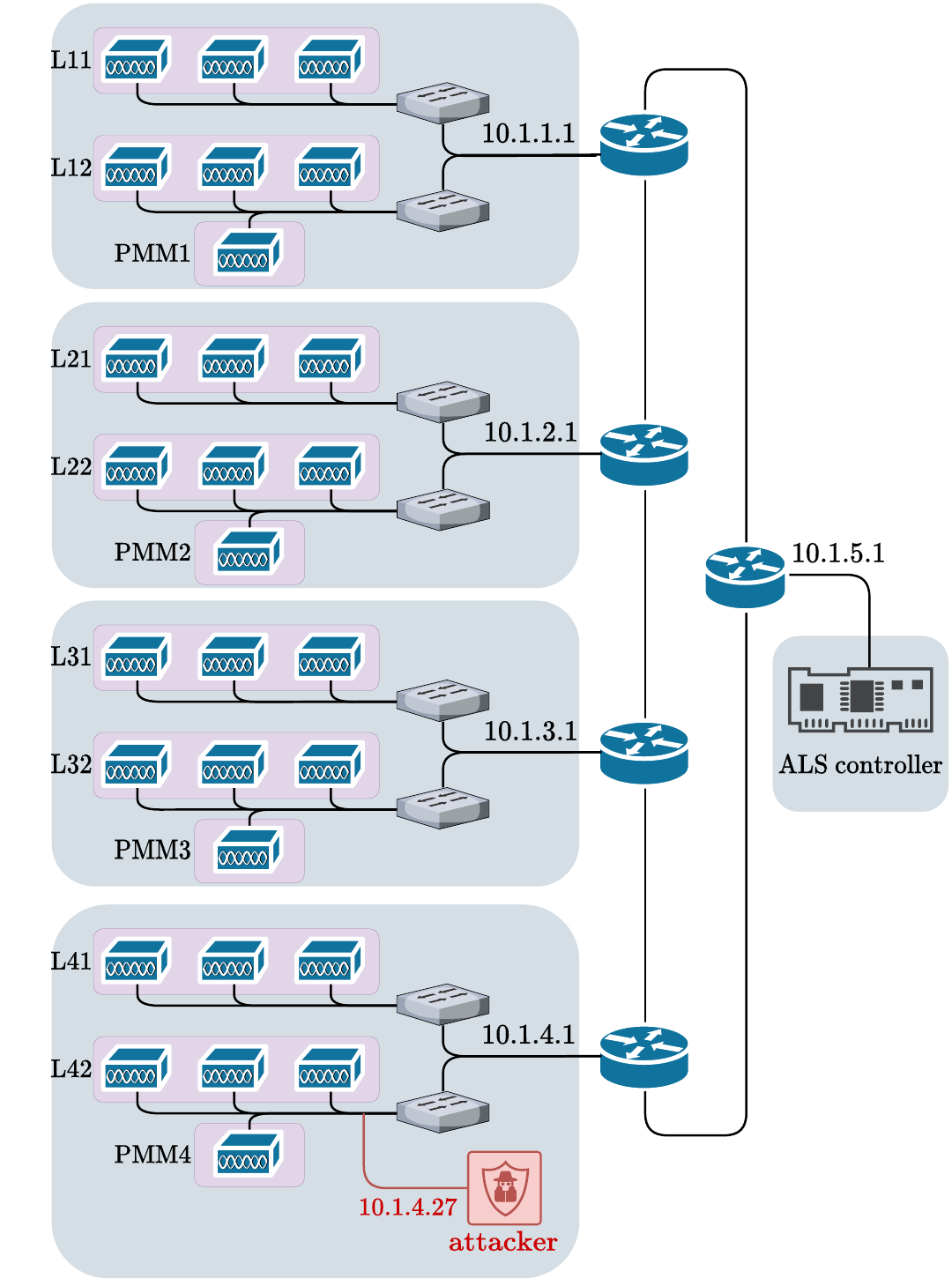}{Communication network.}{fig:network}{0.8}{}

\noindent \rev{Fig. \ref{fig:network} shows the communication network topology of the ship power system, which uses Ethernet cable and consists of five local area networks (LANs). Four LANs are dedicated to the four zones of the MVAC ship power system, and the fifth LAN is for the central controller. The backbone of the network comprises five network routers. Each node in the figure represents a sensor/local controller of ship power devices that is connected to the router through network switches. Each node is assigned a unique IP address, and the communication between nodes is established using the Internet Protocol (IP). For this test, only load centers and PMM devices are included in the communication model. The Modbus TCP protocol is a master-slave protocol that allows a master device to communicate with multiple slave devices. Each device is a Modbus TCP slave, while the ALS controller is the Modbus TCP master. The communication infrastructure is simulated and executed in real-time using $ns3$. The latencies on the networks vary depending on the location of the devices to transfer data from different zones or different LAN networks to the controller (as shown in Fig. \ref{fig:latencies}). Asynchronous data received from devices with different latencies presents challenges to the ALS controller, that are addressed in the design process. The research also presents additional insights into various scenarios, such as changing the rate of data transmission, network congestion, or cyber attacks.}

\rev{The controller node is a hardware node connected to the network via a dedicated Ethernet port, enabling it to interface with external devices or control systems through a HIL setup, which evaluates the performance of the entire complex cyber-physical system. Asynchronous UDP/IP protocol is used to exchange internal signals with external systems. The other nodes are virtual nodes realized by Docker containers, which provide connectivity to devices in the real-time simulated power system. This approach is not limited by communication hardware, resulting in increased system flexibility and adaptability. Different types of communication problems or cyber attacks can be realistically emulated. In Fig. \ref{fig:network}, the red node in the Zone 4 LAN represents an attacker that can launch a cyber attack on the communications network. This node is connected to a container and from this container, a DoS or MITM attack can be generated, making the SCPS suffer from a real attack. The detailed analysis is presented in the next subsection.}

\TIAFigs{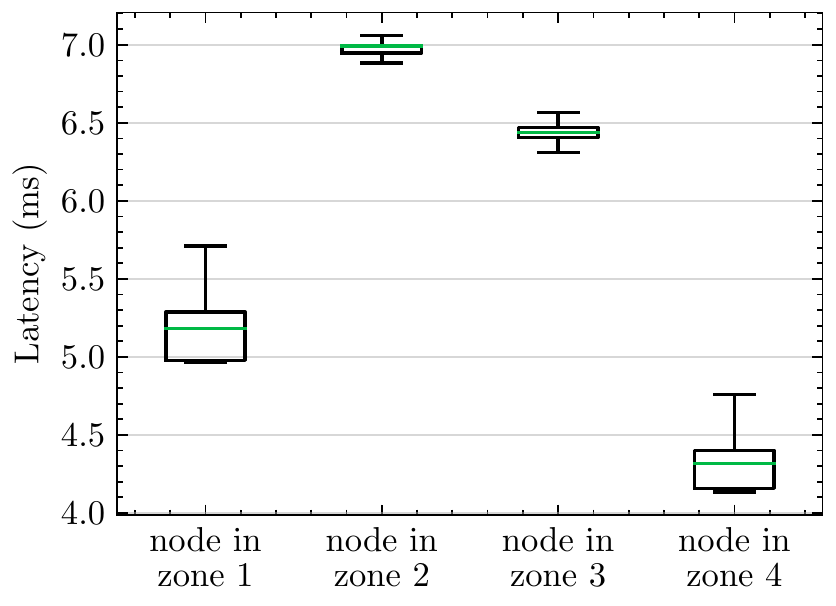}{Communication latencies in different zones.}{fig:latencies}{0.9}{}


\subsection{ALS Controller Performance under Additional Communication Delay}
\label{comdelay2}
\subsubsection{Communication testing}
\rev{In this section, we investigate the impact of communication delay on the performance of the ship's power system under the ALS controller. The issues of communication congestion and packet loss are considered and tested in the Cy-HIL setup.}
\begin{itemize}
	\item \rev{Communication congestion: The changes in network throughput can be observed by varying the rate at which measurement data is sent from devices to the central controller. High network throughput can lead to congestion, which can cause a number of problems, including increased latency, packet loss, and reduced overall network performance. When network traffic is high, the available bandwidth can quickly become saturated, causing delays in sending and receiving data. With all data flowing to the ALS controller, network bottlenecks can occur, exacerbating congestion and further reducing network efficiency. The network throughput at various transmission rates is depicted in Fig. \ref{fig:throughput-congestion}. When measurement signals are sent every 1ms, the average throughput is higher than when they are sent every 30ms. However, this increased rate of transmission can lead to network congestion.}

	      \begin{figure}[t]
		      \centering
		      \subfloat[\label{fig:throughput1}]{%
			      \includegraphics[width=0.9\linewidth]{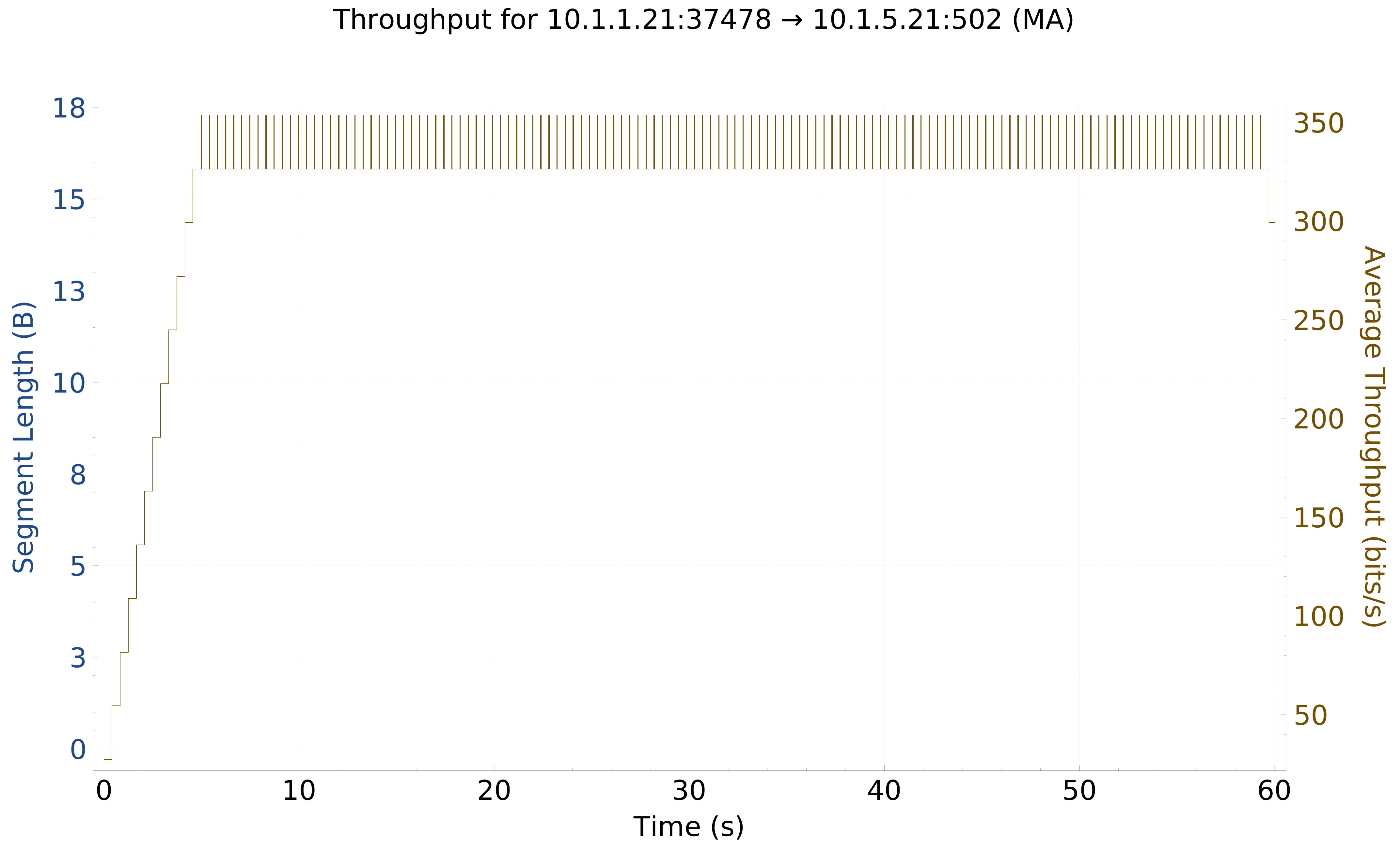}}
		      \\
		      \subfloat[\label{fig:throughput30}]{%
			      \includegraphics[width=0.9\linewidth]{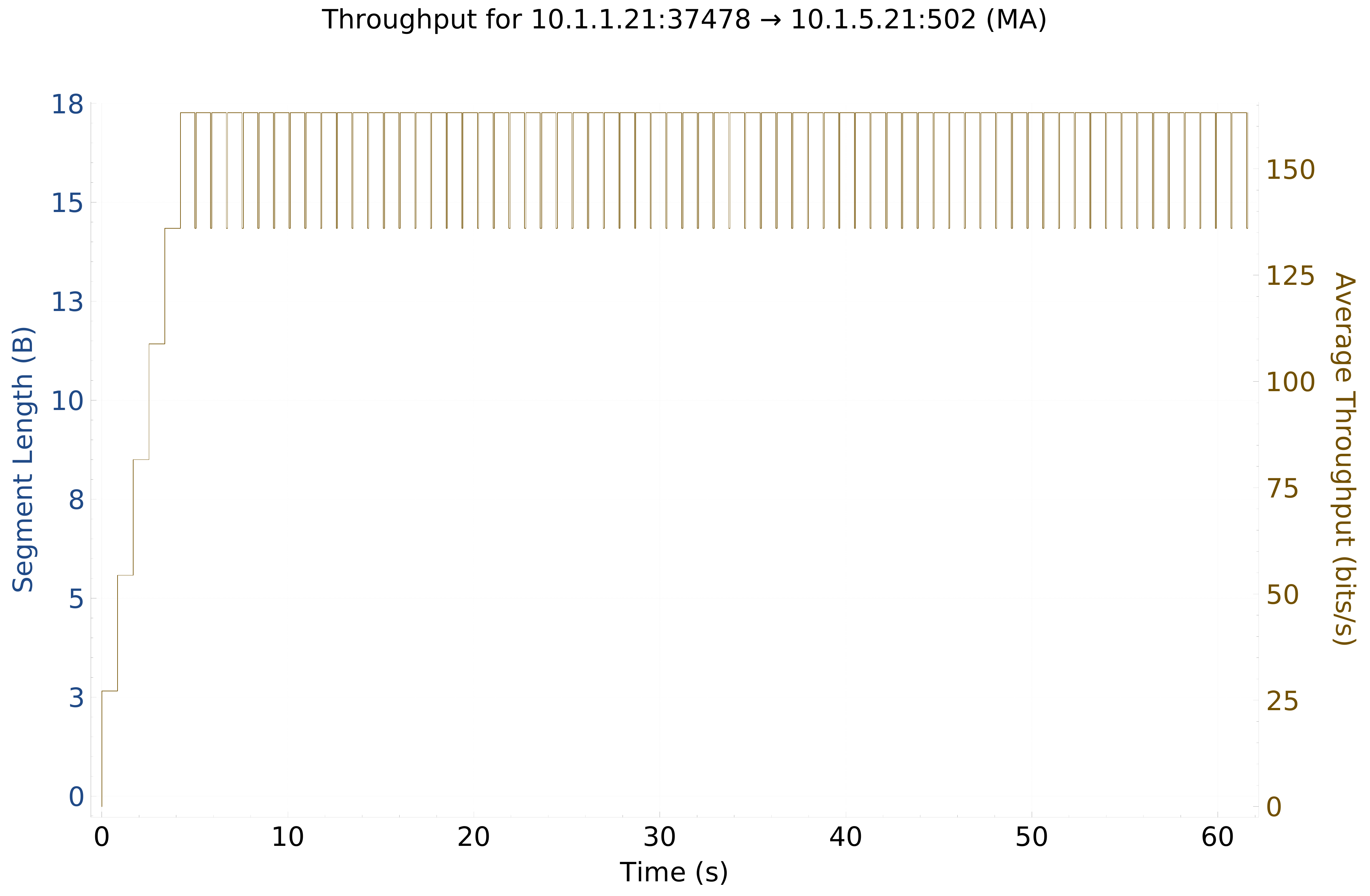}}
		      \caption{\rev{Network throughput (a) messages are sent every 1ms and (b) messages are sent every 30ms.}}
		      \label{fig:throughput-congestion}
	      \end{figure}

	\item \rev{Packet loss: Packet loss occurs when one or more data packets fail to reach their destination due to errors or network congestion. The use of Docker containers as interfaces between cyber and physical systems can be facilitated by the $pumba$ tool \cite{Ledenev} to implement communication testing. $pumba$ emulates the problem of packet loss when transmitting ModbusTCP messages over the network. The phenomenon is captured in $Wireshark$ \cite{10.5555/1202316} as shown in Fig. \ref{fig:packetloss}. The red rows indicate dropped packets that can not reach the destination.}
	      \TIAFigs{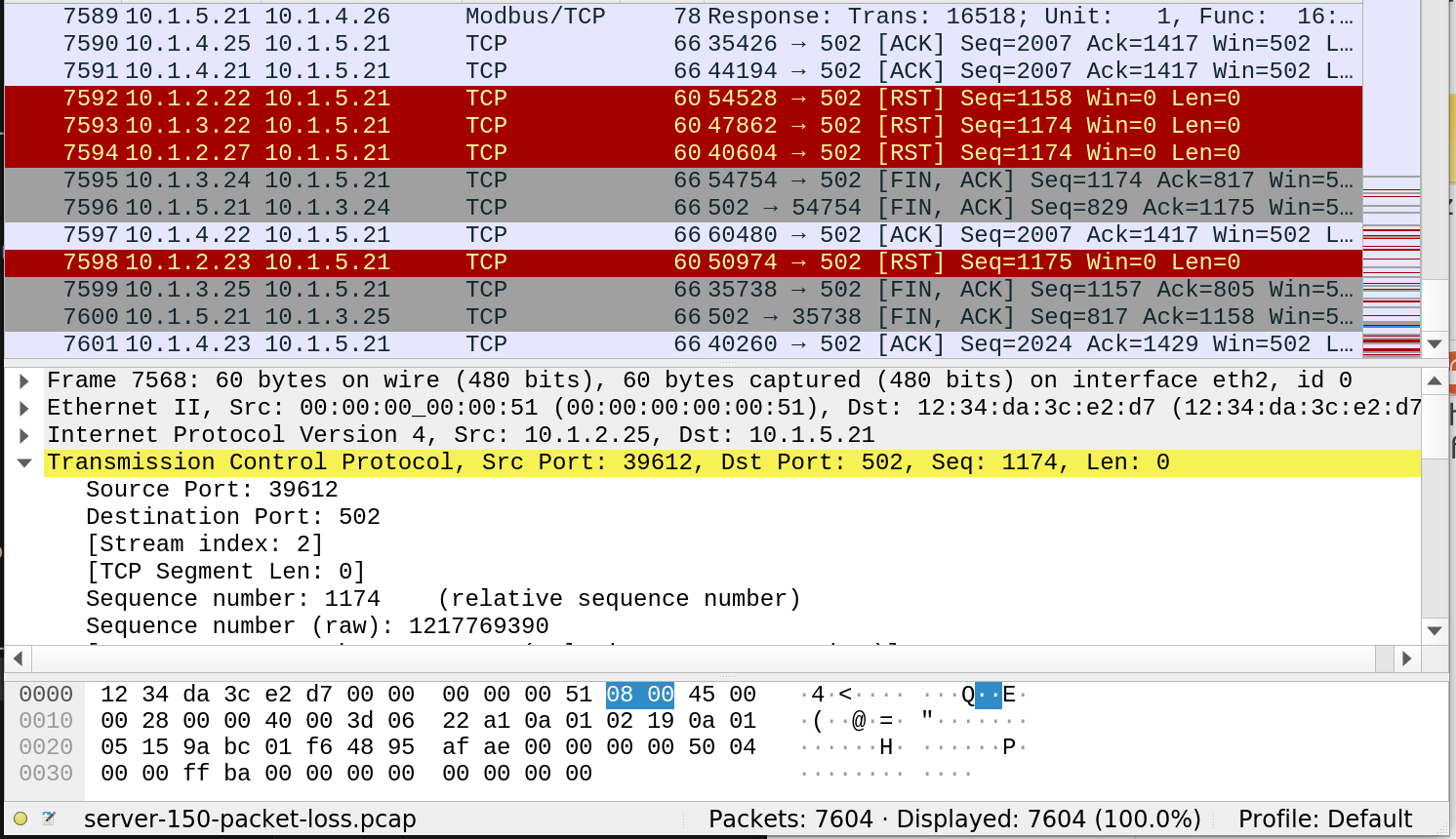}{\rev{Data packets in the network with packet loss issue.}}{fig:packetloss}{0.16}{}
\end{itemize}
\subsubsection{ALS controller performance}
The performance of the ALS controller is evaluated in both synchronous and asynchronous modes with additional latencies to verify the platform's capability to handle time-critical and non-time-critical scenarios and uncover possible communication-related problems. The primary aim is to command the operational status of eight groups of load centers, namely LC11, LC12, $\dots$, LC41, LC42, as illustrated in Fig. \ref{fig:cy_hil}. Each group of load centers has a combined power rating of 1 MW and consists of three types of loads: vital loads (0.65 MW), semi-vital loads (0.1 MW), and non-vital loads (0.25 MW). Additionally, the ALS controller manages four propulsion motor modules (PMM), with each module having a power rating of 15 MW.

\begin{figure}[t]
	\centering
	\subfloat[\label{6a}]{%
		\includegraphics[width=0.9\linewidth]{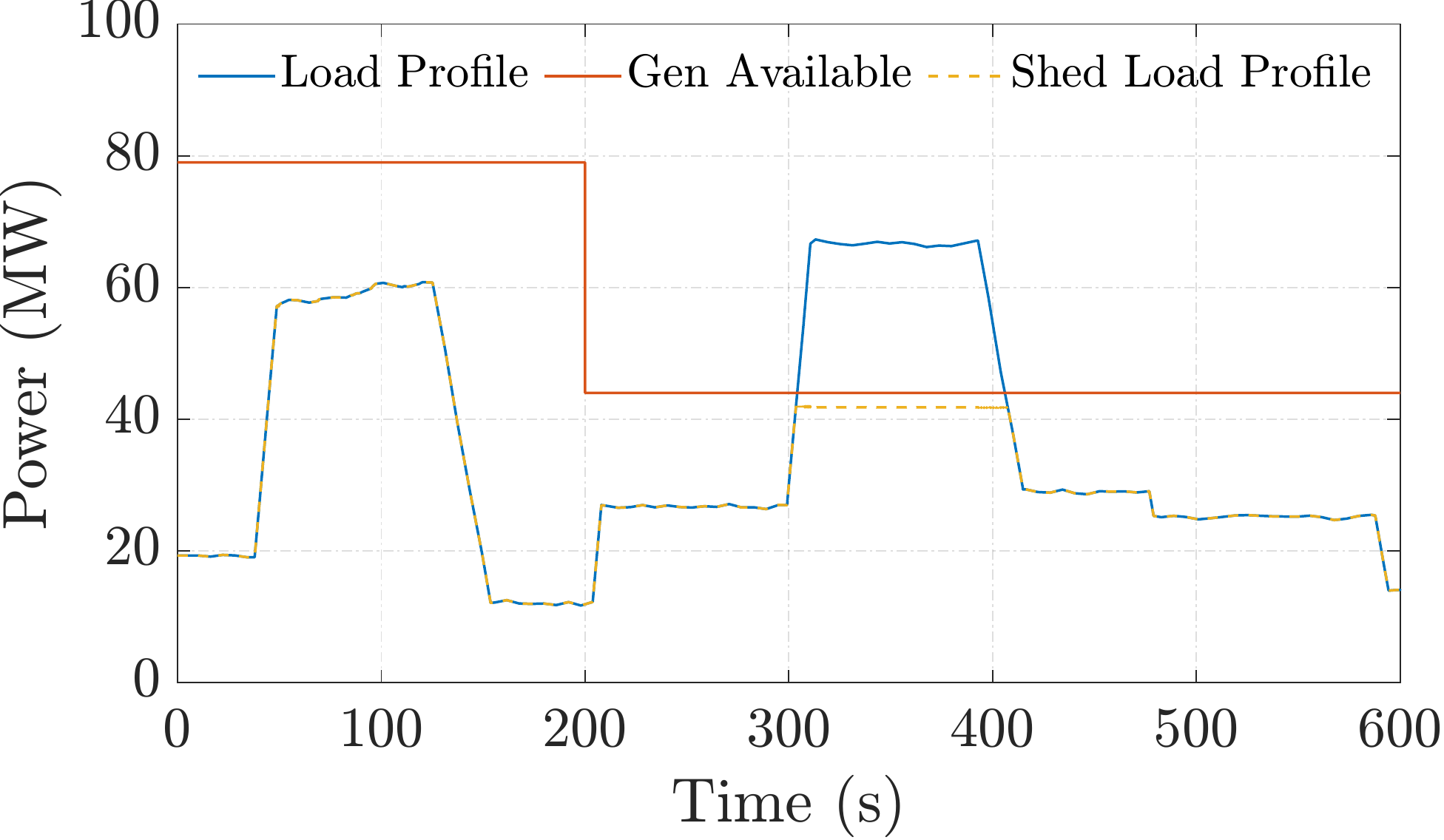}}
	\\
	\subfloat[\label{6b}]{%
		\includegraphics[width=0.9\linewidth]{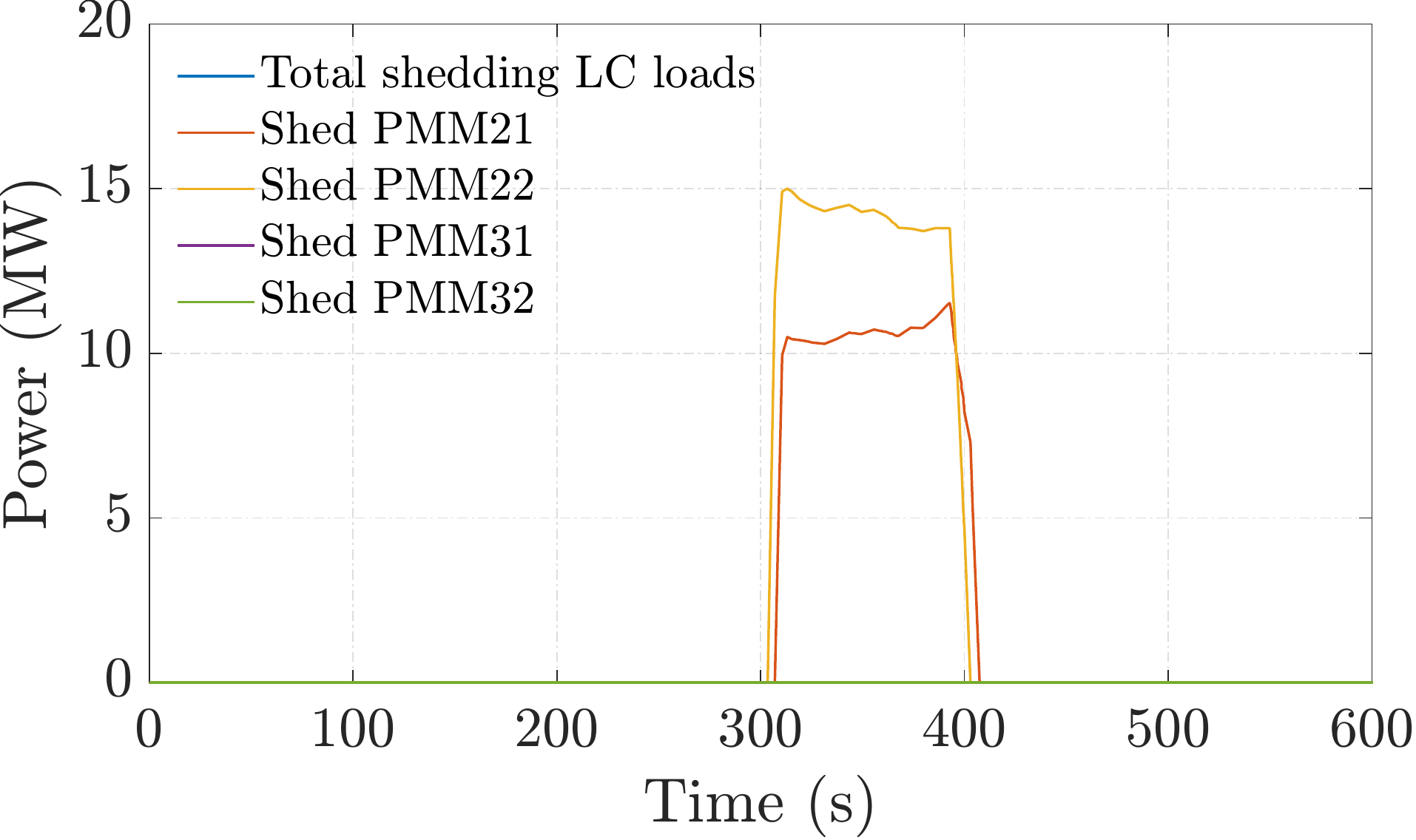}}
	\caption{Physical performance in synchronous mode (a) Ship power system measurements and (b) ALS controller outputs.}
	\label{fig1}
\end{figure}

Fig. (\ref{6a}) and (\ref{6b}) illustrate the response of the ALS controller in a synchronous communication scenario following an over-power event that occurred from $300$s to $400$s, triggered by the unavailability of one of the main generators (rated at 35 MW) at $t = 200$s. In this case, PMM21 and PMM22 are shed to ensure that the total served power is always lower than the total generator power. Note that all load centers are left unshed to maximize load operability and reduce frequent load switching.

\begin{figure}[t]
	\centering
	\subfloat[\label{7a}]{%
		\includegraphics[width=0.9\linewidth]{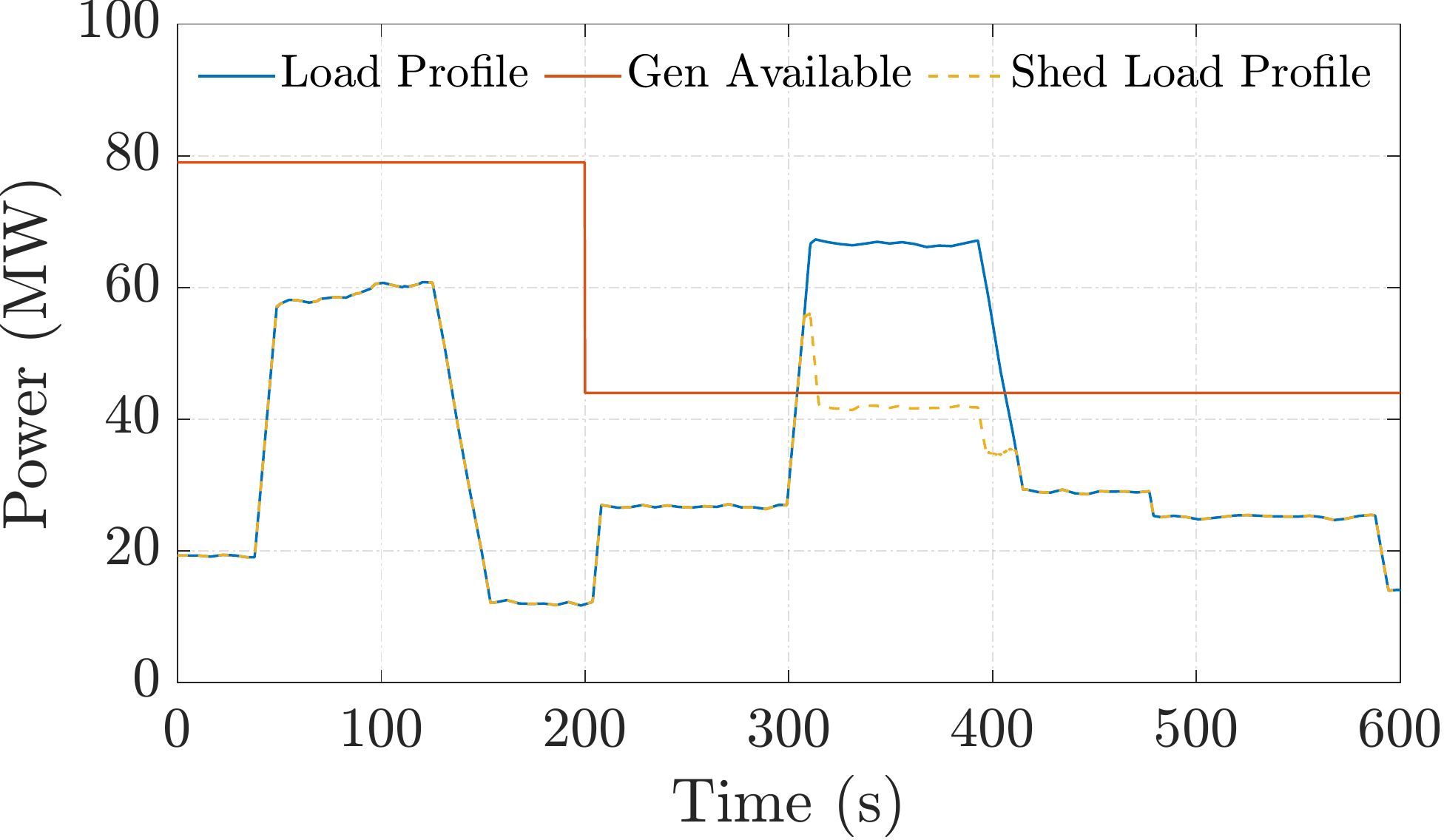}}
	\\
	\subfloat[\label{7b}]{%
		\includegraphics[width=0.9\linewidth]{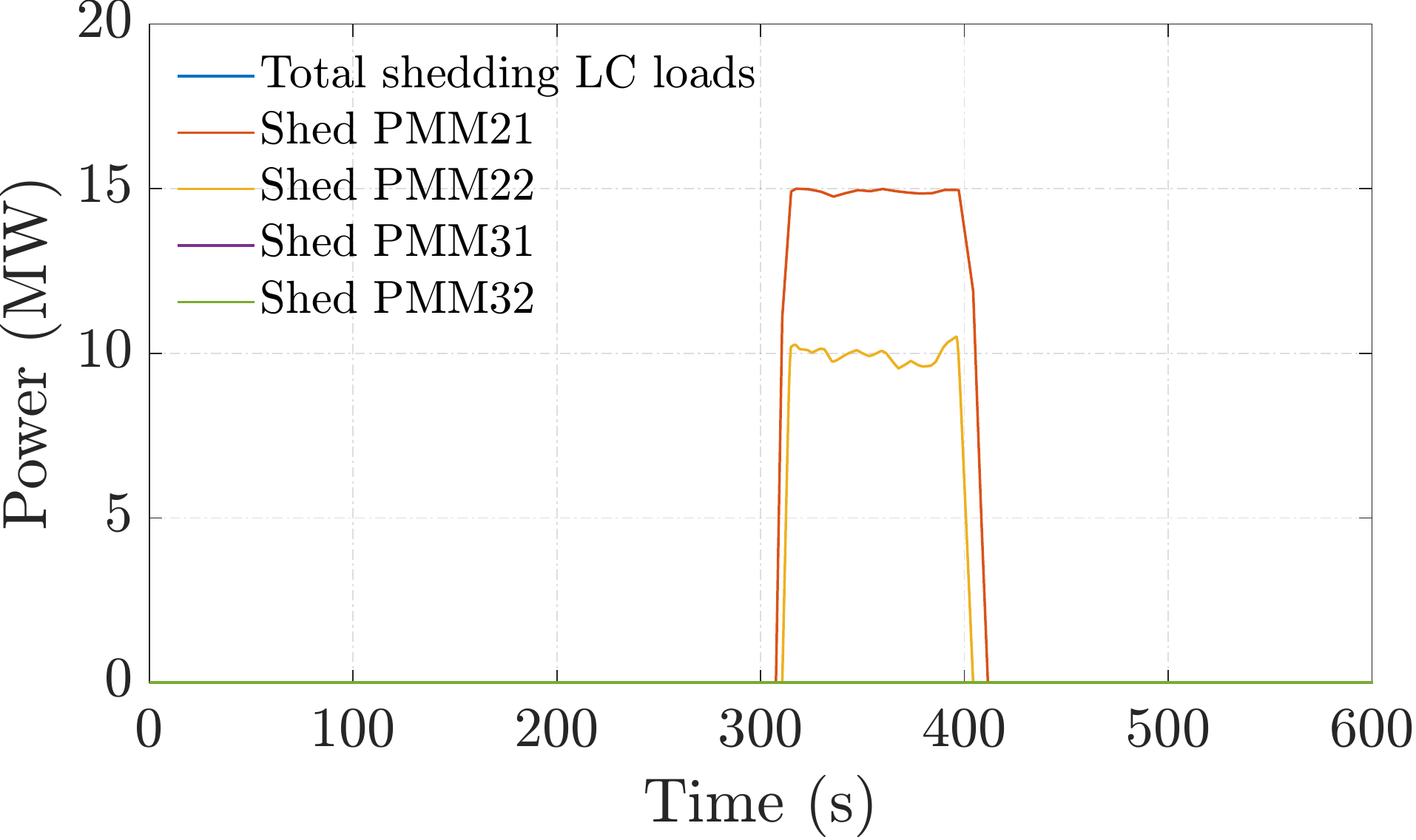}}
	\caption{Physical performance in asynchronous mode (a) Ship power system measurements and (b) ALS controller outputs.}
	\label{fig2}
\end{figure}

The next simulation setup remains the same, except for the addition of  communication delay time for each measurement signal sent from $OPAL-RT$. The communication delay is caused by the network congestion presented in the previous subsection. As shown in Figs. (\ref{7a}) and (\ref{7b}), due to the asynchronous communication, the controller failed to shed loads between $309$s and $314$s and shed excessive loads between $393$s and $406$s. Moreover, the set of shed loads was also changed to PMM21 and PMM31 instead of PMM21 and PMM22.

\subsection{ALS Controller Performance under Cyber Attacks}
\label {cyberattack}
\noindent The test cases in this section are designed to examine the performance of the ALS controller in the face of cyber-attacks. The investigation assumes that an attacker can gain access to the communications network and issue malicious commands to the controller. The attacker is located in LAN zone 4, which is connected to the controller via routers. To realistically simulate the attack, an additional Docker container is created within the Cyber-HIL platform using open-source tools for support. Packet data is captured from a live network using a network protocol analyzer, specifically $tshark$, and then visualized using $Wireshark$.

\subsubsection{Denial of Service (DoS) attack}

A DoS attack is a type of cyber attack designed to disrupt the normal functioning of a target system, network, or service by overwhelming it with traffic or requests. The attack aims to make the targeted resource unavailable to legitimate users, resulting in a service outage or downtime. The attack can also be launched from a single source (single-source DoS) or multiple sources (distributed DoS).

In this test, the attacker employs the $hping3$ tool \cite{Sanfilippo_2022} to launch a DoS attack by flooding the target with traffic. $hping3$ is a command-line tool that generates DoS attacks, allowing the user to send TCP, UDP, and ICMP packets to a target host, with the capability to specify the source IP address and port number. Since Modbus TCP is utilized in the control system, the attacker targets port 502/TCP on the controller computer, which is commonly used for Modbus TCP communication. The attack is launched between 295s and 370s, causing a large number of packets to be sent to the target, overwhelming the controller and causing it to crash. Fig. \ref{fig:dos}, Fig. \ref{fig:round_trip}, and Fig. \ref{fig:throughput} are obtained from $Wireshark$ to examine the network performance. Fig. \ref{fig:dos} depicts the total packets and the number of packets from the attacker, revealing a surge in data packets during the attack. This leads to a significant increase in the round trip time (RTT) and a decrease in the throughput, as shown in Fig. \ref{fig:round_trip} and Fig. \ref{fig:throughput}. The RTT refers to the time taken to send a packet and receive an acknowledgment, while the throughput measures the rate of successful message delivery over a communication channel. The results indicate that the network is congested during the attack, causing a delay in the transmission of data packets and a decrease in the throughput. The controller becomes unable to receive commands from the operator and fails to shed loads, resulting in the total served power exceeding the total generator power and potentially leading to a blackout.

\begin{figure}[t]
	\centering
	\subfloat[\label{dos_result_1}]{%
		\includegraphics[width=0.9\linewidth]{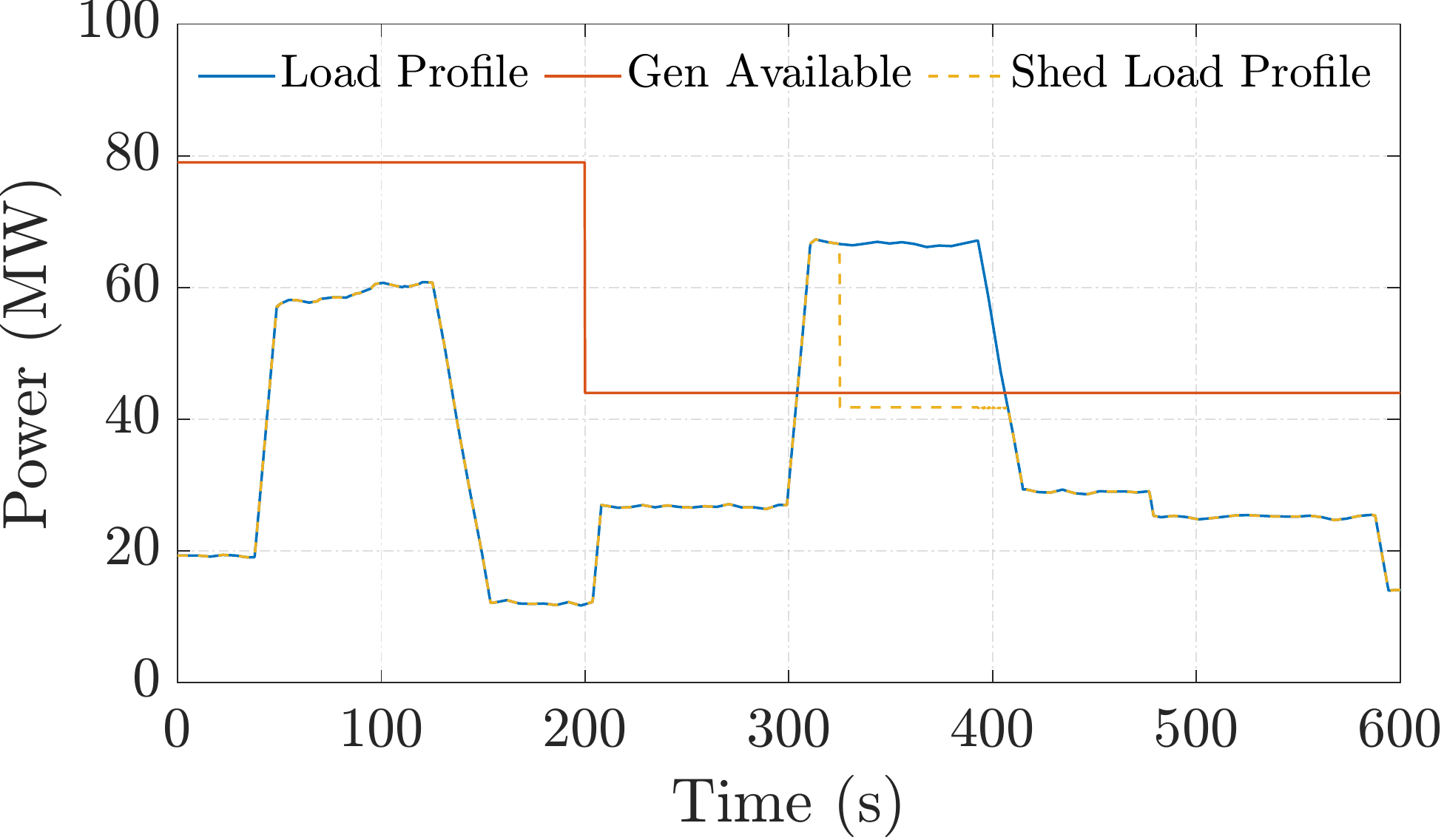}}
	\\
	\subfloat[\label{dos_result_2}]{%
		\includegraphics[width=0.9\linewidth]{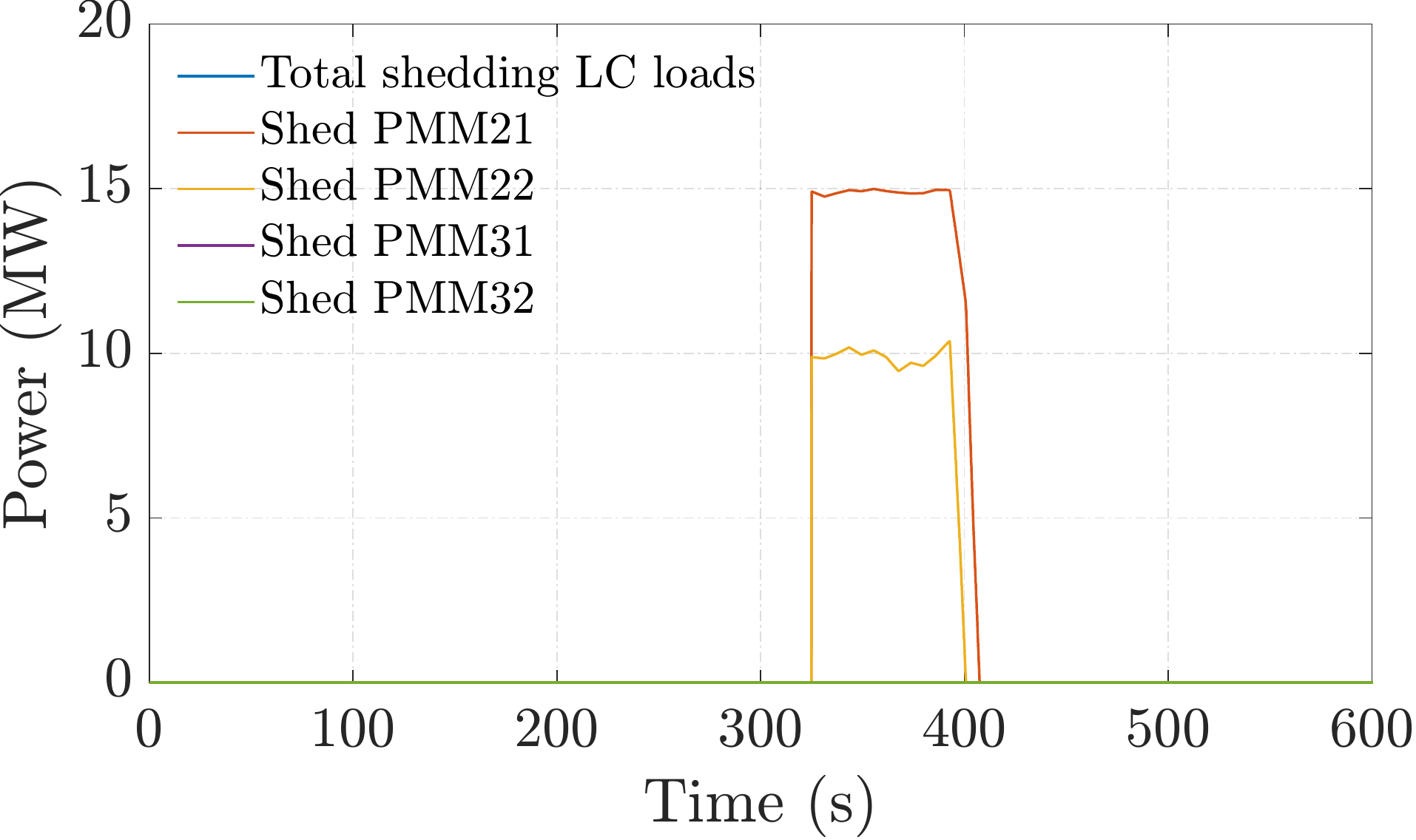}}
	\caption{Physical performance under DOS attack (a) Ship power system measurements and (b) ALS controller outputs.}
	\label{dos}
\end{figure}

Fig. \ref{dos} demonstrates the performance of the controller. It is evident from the plot that the controller function is obstructed during the attack, leading to a failure in shedding loads as necessary when the load power surpasses the generators' capacity. This results in an imbalance between generator output and load demand from 300s to 325s, with total supplied power exceeding total generator output. After the attack is over, the controller resumes its function and the loads are shed to ensure that the total served power is always less than the total generator power. The results indicate that the controller is unable to shed loads during the attack, which can harmfully impact the generators due to overload.

\TIAFigs{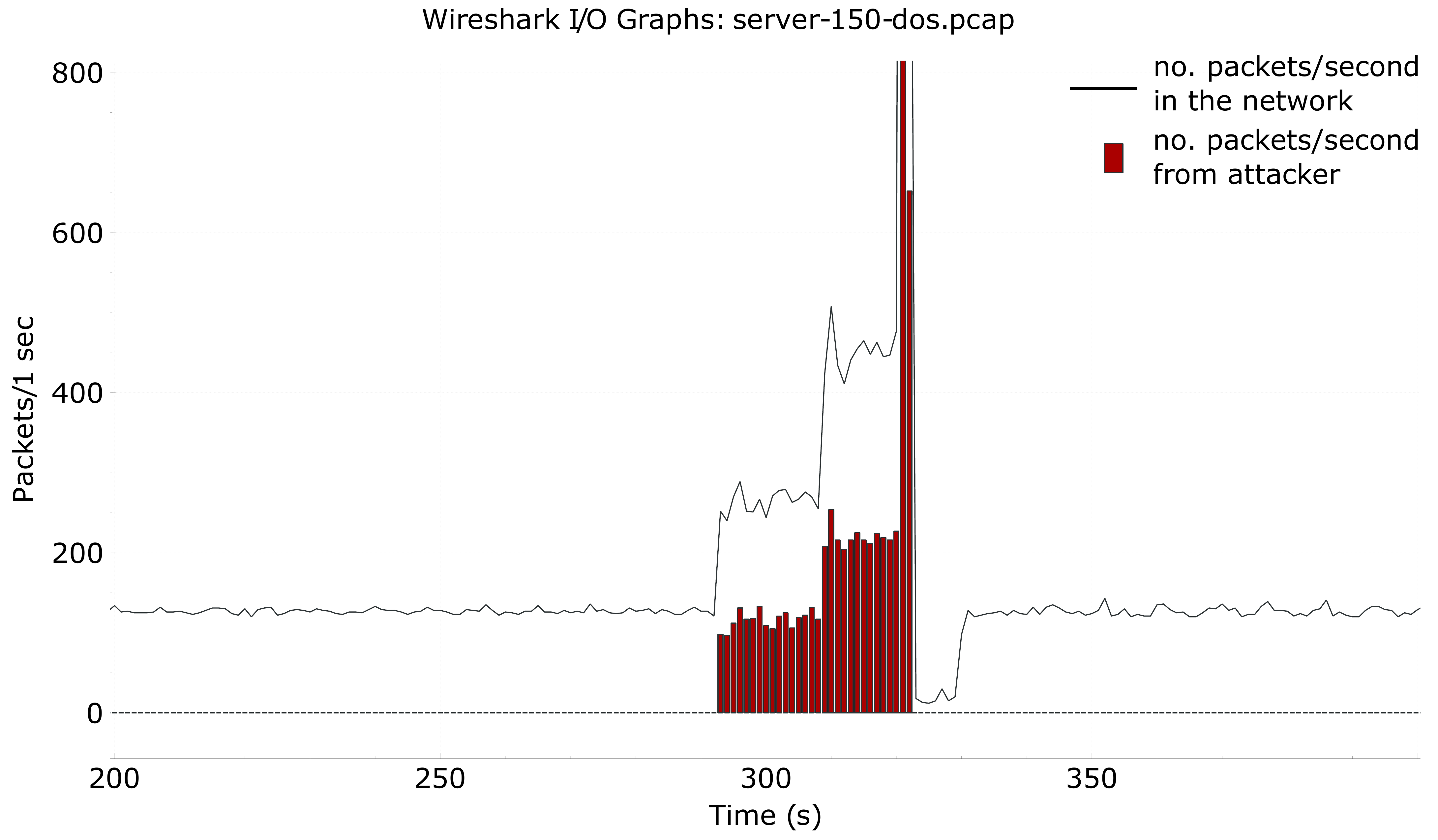}{Data packets in the network in the DOS attack scenario.}{fig:dos}{0.2}{}

\TIAFigs{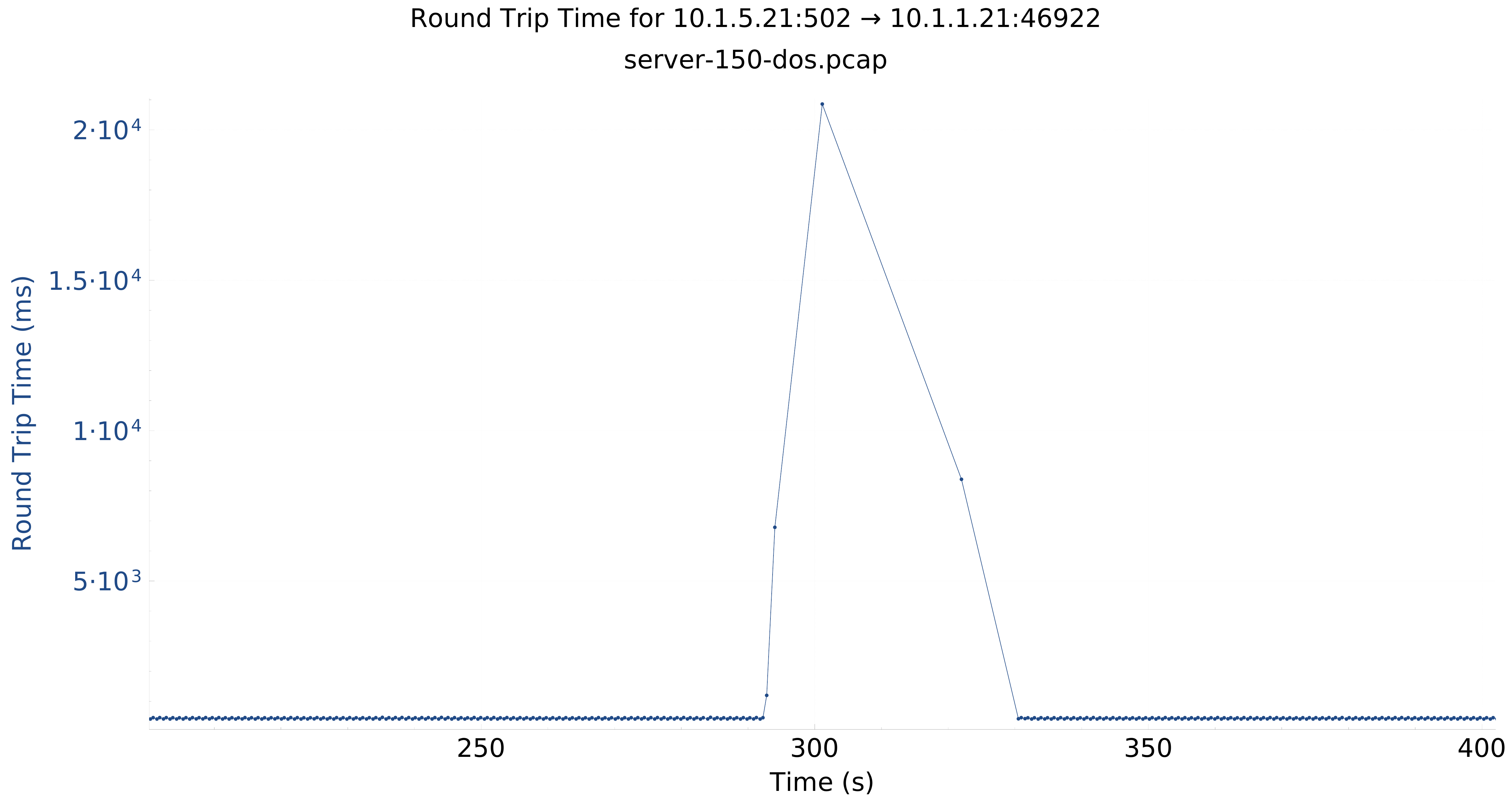}{Round trip time (RTT) of Modbus TCP packet between the ALS controller and a device in LAN zone 1.}{fig:round_trip}{0.17}{}

\TIAFigs{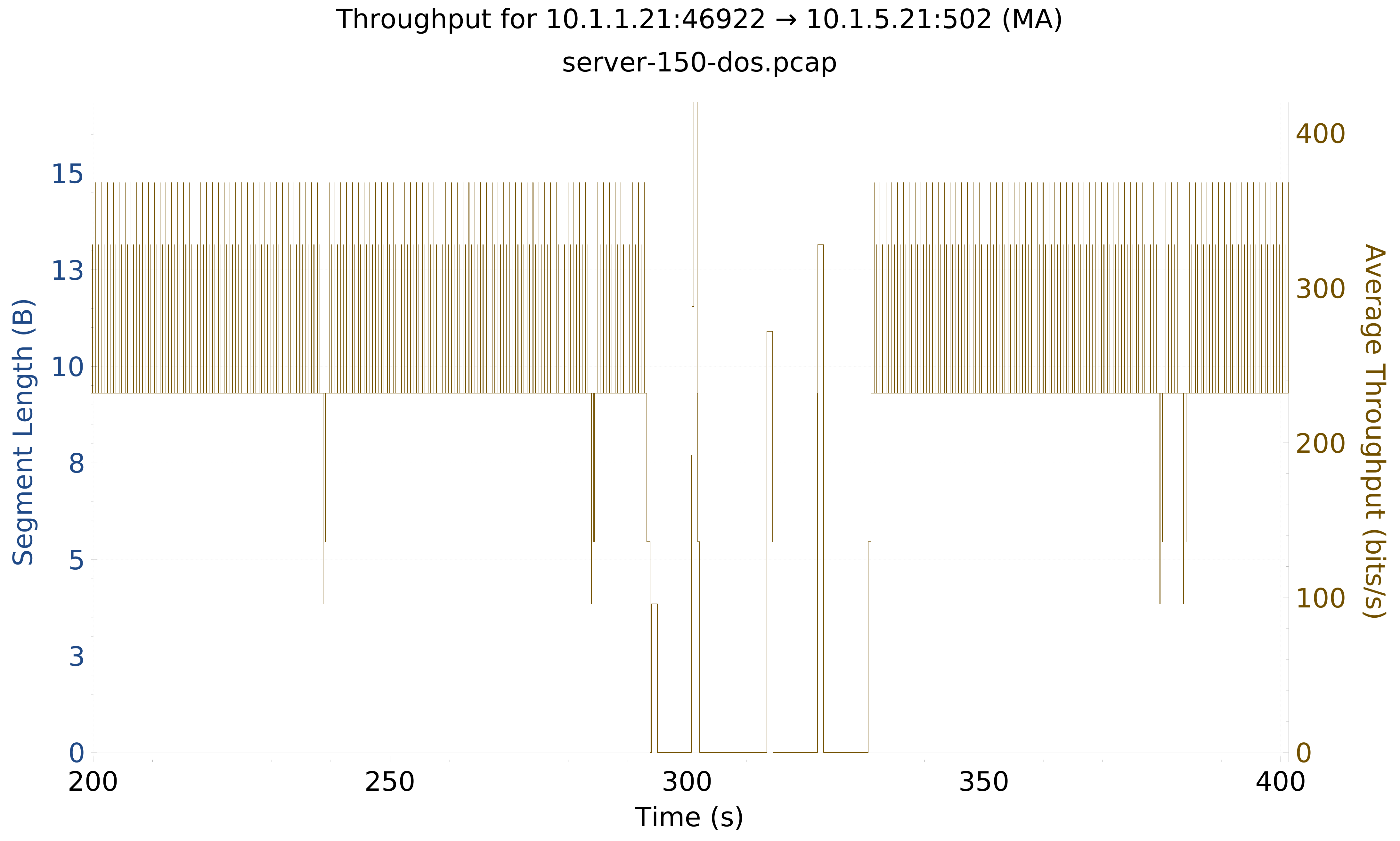}{Throughput for the messages between the ALS controller and a device in LAN zone 1.}{fig:throughput}{0.17}{}


\subsubsection{ALS controller performance under Man-in-the-Middle attacks}

\noindent This section describes a Man-in-the-Middle (MITM) attack, performed by the same attacker as in the previous section. This time, $Scapy$ tool \cite{8903954} is utilized to conduct the attack. The MITM attack involves intercepting communication between the Modbus TCP server/ALS controller and a Modbus TCP client/device, allowing the attacker to eavesdrop on conversations or modify the messages exchanged. As shown in Fig. \ref{fig:mimt}, the measurement signals from the PMM4 are routed through the attacker's computer instead of reaching the controller directly. To achieve this, the attack follows a three-step process. First, the attacker poisons the Address Resolution Protocol (ARP) tables on the master and slave devices, tricking the network into trusting the predefined route for communication. Second, as all traffic between the two devices passes through the attacker's machine, specific packets are captured and modified using $Scapy$'s sniffing function, which applies a given function to any packet that meets the specified criteria. Finally, the ARP tables are purged to remove the attacker's code from the communication, making post-attack analysis more difficult.

\begin{figure}[t]
	\centering
	\subfloat[\label{mitm_result_1}]{%
		\includegraphics[width=0.9\linewidth]{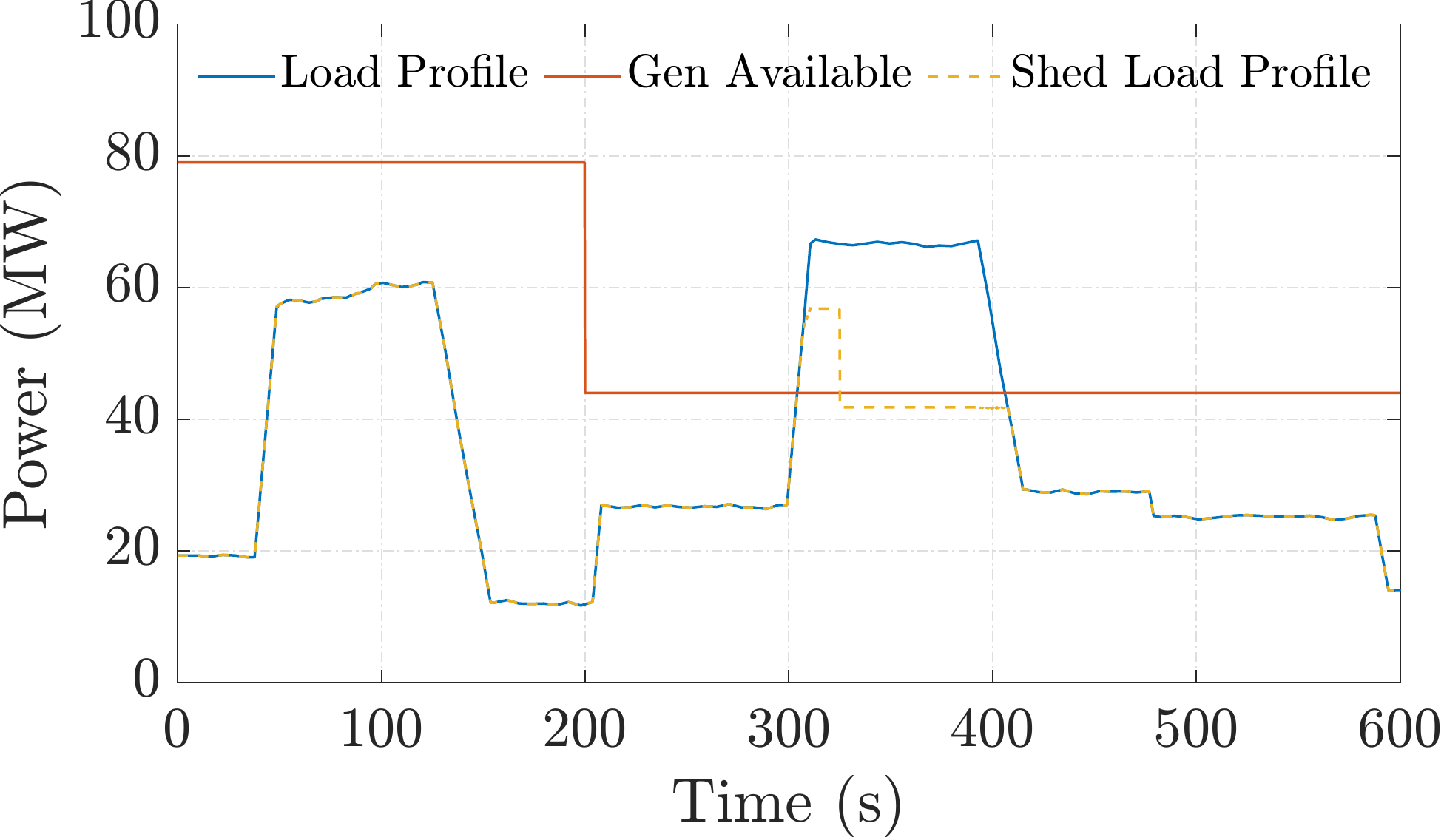}}
	\\
	\subfloat[\label{mitm_result_2}]{%
		\includegraphics[width=0.9\linewidth]{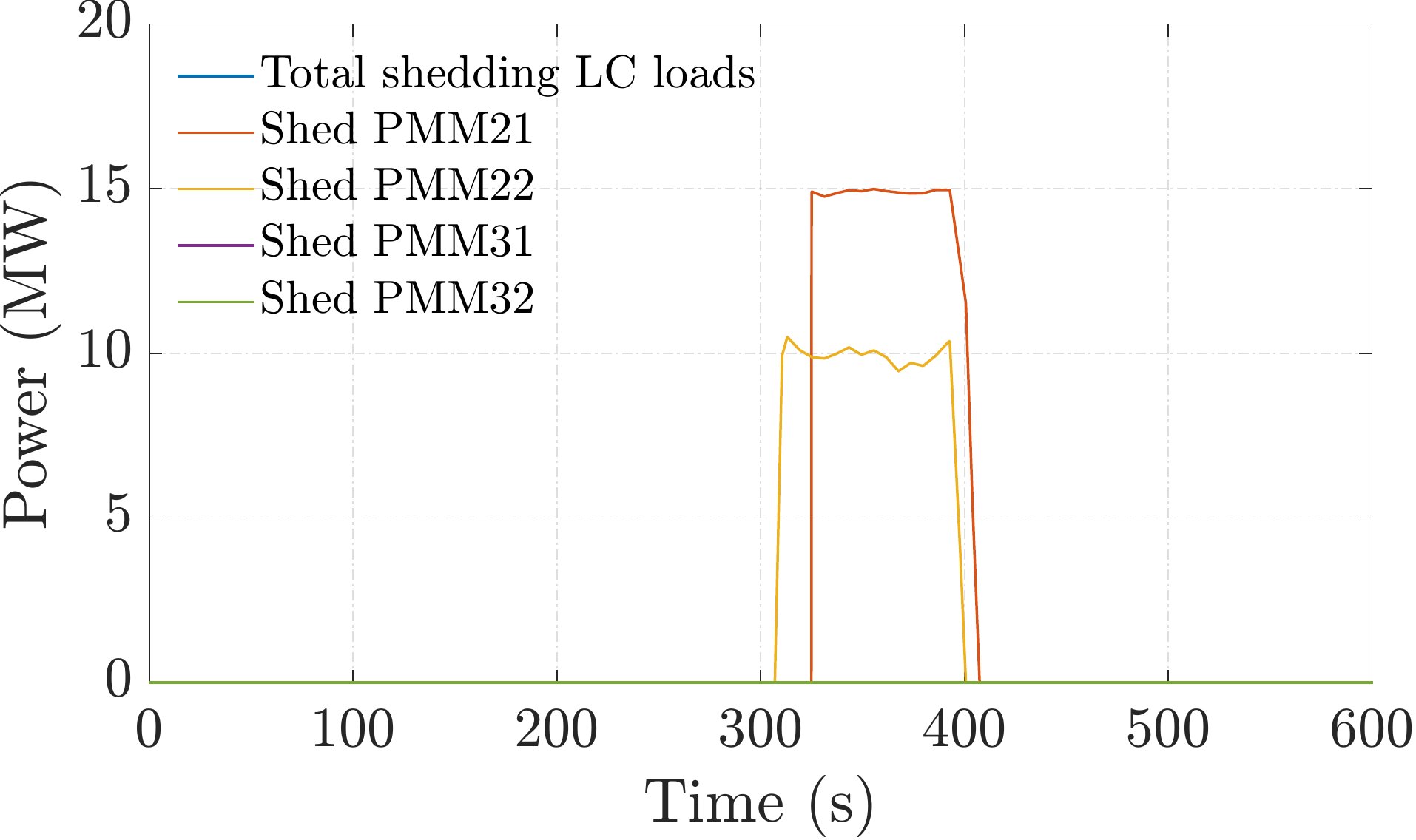}}
	\caption{Physical performance under MITM attack (a) Ship power system measurements and (b) ALS controller outputs.}
	\label{mitm_result}
\end{figure}

\rev{During the test, a MITM attack occurred between 295s and 325s, intercepting Modbus TCP packets sent from the PMM4 to the controller. The intercepted data was then altered to set a value of 0 during this timeframe. The impact of this attack on the control system is depicted in Fig. \ref{mitm_result}, revealing that the controller provided inappropriate control signals to the PMMs, resulting in the total delivered power exceeding the total generator power. This misbehavior can be attributed to the controller receiving inaccurate information about the ship's power system, as the total load power was manipulated by the MITM attacker, causing it to appear lower than the actual value.} \rev{Upon completion of the attack, the controller gains an actual view of the system and efficiently sheds loads to maintain the total served power below the total generator power. Nevertheless, the outcomes reveal that the controller encountered difficulties in shedding loads during the attack, posing a potential risk of a blackout occurrence.}

\TIAFigs{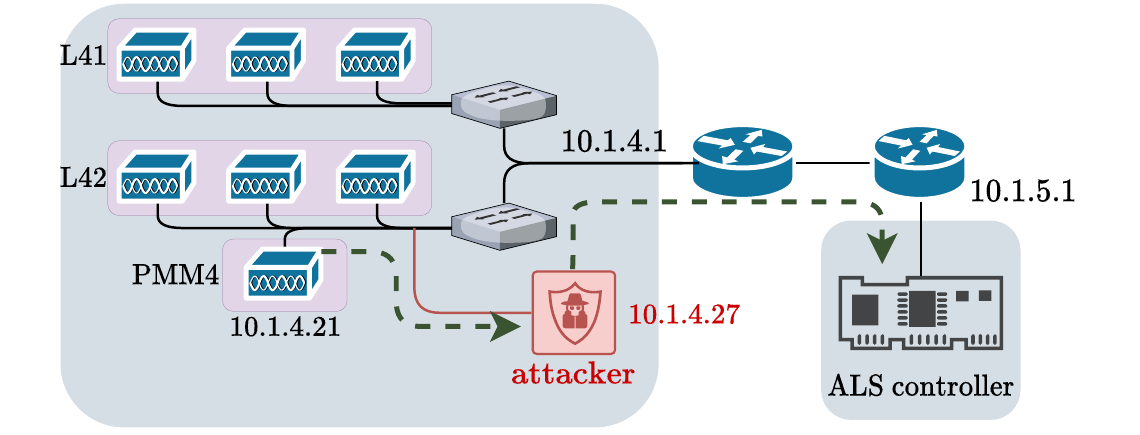}{Data flow from PMM4 to ALS controller under the MITM attack scenario.}{fig:mimt}{0.75}{}

\section{Conclusion}
\noindent
\rev{This paper presents a Cyber-HIL approach for investigating control systems in SCPS. The proposed approach integrates hardware, software, and network components, creating a realistic and comprehensive simulation environment. Experimental results demonstrate the effectiveness of the Cyber-HIL approach in detecting and mitigating cyber-attacks, including MITM and DOS attacks, in SCPS. The Cyber-HIL approach also serves as a valuable platform for assessing SCPS resilience and evaluating control algorithm performance. Future research opportunities include applying the Cyber-HIL approach to other Cyber-Physical Systems, such as aviation, transportation, and industrial systems. Enhancing the cybersecurity and reliability of CPS, this approach contributes significantly to safe and efficient operations in various domains. In the future, we aim to investigate the proposed approach in more complex ship power systems with multiple controllers, expanding the communication network, and employing threat modeling to comprehensively evaluate the cyber security of SCPS. These efforts advance CPS robustness and security, meeting critical demands of safe and dependable operations.}




\bibliographystyle{IEEEtran}
\bibliography{ref, zotero_refs}

\begin{thebibliography}{10}
\providecommand{\url}[1]{#1}
\csname url@samestyle\endcsname
\providecommand{\newblock}{\relax}
\providecommand{\bibinfo}[2]{#2}
\providecommand{\BIBentrySTDinterwordspacing}{\spaceskip=0pt\relax}
\providecommand{\BIBentryALTinterwordstretchfactor}{4}
\providecommand{\BIBentryALTinterwordspacing}{\spaceskip=\fontdimen2\font plus
\BIBentryALTinterwordstretchfactor\fontdimen3\font minus \fontdimen4\font\relax}
\providecommand{\BIBforeignlanguage}[2]{{%
\expandafter\ifx\csname l@#1\endcsname\relax
\typeout{** WARNING: IEEEtran.bst: No hyphenation pattern has been}%
\typeout{** loaded for the language `#1'. Using the pattern for}%
\typeout{** the default language instead.}%
\else
\language=\csname l@#1\endcsname
\fi
#2}}
\providecommand{\BIBdecl}{\relax}
\BIBdecl

\bibitem{Progoulakis2023}
I.~Progoulakis, N.~Nikitakos, D.~Dalaklis, A.~Christodoulou, A.~Dalaklis, and R.~Yaacob, ``Digitalization and cyber physical security aspects in maritime transportation and port infrastructure,'' in \emph{Smart Ports and Robotic Systems : {{Navigating}} the Waves of Techno-Regulation and Governance}.\hskip 1em plus 0.5em minus 0.4em\relax {Cham}: {Springer International Publishing}, 2023, pp. 227--248.

\bibitem{jmse9121384}
I.~Progoulakis, P.~Rohmeyer, and N.~Nikitakos, ``Cyber {{Physical Systems Security}} for {{Maritime Assets}},'' \emph{Journal of Marine Science and Engineering}, vol.~9, no.~12, p. 1384, Dec. 2021.

\bibitem{GieringDyck+2021+1081+1095}
J.-E. Giering and A.~Dyck, ``Maritime {{Digital Twin}} architecture: {{A}} concept for holistic {{Digital Twin}} application for shipbuilding and shipping,'' \emph{at - Automatisierungstechnik}, vol.~69, no.~12, pp. 1081--1095, Dec. 2021.

\bibitem{markle2019naval}
S.~Markle and T.~Moore, ``Naval {{Power}} and {{Energy Systems}}: {{Technology Development Roadmap}},'' {US Navy Sea Systems Command}, Tech. Rep., 2019.

\bibitem{KARAHALIOS20181}
H.~Karahalios, ``The severity of shipboard communication failures in maritime emergencies: {{A}} risk management approach,'' \emph{International Journal of Disaster Risk Reduction}, vol.~28, pp. 1--9, Jun. 2018.

\bibitem{9329185}
G.~Zhang, J.~Shi, S.~Huang, J.~Wang, and H.~Jiang, ``A {{Cascading Failure Model Considering Operation Characteristics}} of the {{Communication Layer}},'' \emph{IEEE Access}, vol.~9, pp. 9493--9504, 2021.

\bibitem{boyes2017code}
H.~Boyes and R.~Isbell, \emph{Code of Practice: Cyber Security for Ships}.\hskip 1em plus 0.5em minus 0.4em\relax {Institution of Engineering and Technology}, 2017.

\bibitem{YAACOUB2020103201}
J.-P.~A. Yaacoub, O.~Salman, H.~N. Noura, N.~Kaaniche, A.~Chehab, and M.~Malli, ``Cyber-physical systems security: {{Limitations}}, issues and future trends,'' \emph{Microprocessors and Microsystems}, vol.~77, p. 103201, Sep. 2020.

\bibitem{9494202}
J.~Choi, D.~Narayanasamy, B.~Ahn, S.~Ahmad, J.~Zeng, and T.~Kim, ``A {{Real-Time Hardware-in-the-Loop}} ({{HIL}}) {{Cybersecurity Testbed}} for {{Power Electronics Devices}} and {{Systems}} in {{Cyber-Physical Environments}},'' in \emph{2021 {{IEEE}} 12th {{International Symposium}} on {{Power Electronics}} for {{Distributed Generation Systems}} ({{PEDG}})}, Jun. 2021, pp. 1--5.

\bibitem{9096292}
Z.~Liu, Q.~Wang, and Y.~Tang, ``Design of a {{Cosimulation Platform With Hardware-in-the-Loop}} for {{Cyber-Attacks}} on {{Cyber-Physical Power Systems}},'' \emph{IEEE Access}, vol.~8, pp. 95\,997--96\,005, 2020.

\bibitem{9184969}
V.~H. Nguyen, T.~L. Nguyen, Q.~T. Tran, Y.~Besanger, and R.~Caire, ``Integration of {{SCADA Services}} and {{Power-Hardware-in-the-Loop Technique}} in {{Cross-Infrastructure Holistic Tests}} of {{Cyber-Physical Energy Systems}},'' \emph{IEEE Transactions on Industry Applications}, vol.~56, no.~6, pp. 7099--7108, Nov. 2020.

\bibitem{9526562}
L.~Faramondi, F.~Flammini, S.~Guarino, and R.~Setola, ``A {{Hardware-in-the-Loop Water Distribution Testbed Dataset}} for {{Cyber-Physical Security Testing}},'' \emph{IEEE Access}, vol.~9, pp. 122\,385--122\,396, 2021.

\bibitem{4906565}
R.~Fang, W.~Jiang, J.~Khan, and R.~Dougal, ``System-level thermal modeling and co-simulation with hybrid power system for future all electric ship,'' in \emph{2009 {{IEEE Electric Ship Technologies Symposium}}}, Apr. 2009, pp. 547--553.

\bibitem{MIC-2020-4-2}
L.~I. Hatledal, R.~Skulstad, G.~Li, A.~Styve, and H.~Zhang, ``Co-simulation as a {{Fundamental Technology}} for {{Twin Ships}},'' \emph{Modeling, Identification and Control: A Norwegian Research Bulletin}, vol.~41, no.~4, pp. 297--311, 2020.

\bibitem{5619971}
M.~Albu, G.~T. Heydt, and S.-C. Cosmescu, ``Versatile platforms for wide area synchronous measurements in power distribution systems,'' in \emph{North {{American Power Symposium}} 2010}, Sep. 2010, pp. 1--7.

\bibitem{9325305}
S.~Jie and Z.~Jian, ``Research {{On Synchronization Technology Of Peer-to-peer Distributed Real-time Database Based On Ship Platform}},'' in \emph{2020 {{International Symposium}} on {{Computer Engineering}} and {{Intelligent Communications}} ({{ISCEIC}})}, Aug. 2020, pp. 70--75.

\bibitem{9046860}
E.~Dushku, M.~M. Rabbani, M.~Conti, L.~V. Mancini, and S.~Ranise, ``{{SARA}}: {{Secure Asynchronous Remote Attestation}} for {{IoT Systems}},'' \emph{IEEE Transactions on Information Forensics and Security}, vol.~15, pp. 3123--3136, 2020.

\bibitem{8412096}
J.~{Al-Jaroodi} and N.~Mohamed, ``{{PsCPS}}: {{A Distributed Platform}} for {{Cloud}} and {{Fog Integrated Smart Cyber-Physical Systems}},'' \emph{IEEE Access}, vol.~6, pp. 41\,432--41\,449, 2018.

\bibitem{8433151}
M.~S. Lund, J.~E. Gulland, O.~S. Hareide, v.~J{\o}sok, and K.~O.~C. Weum, ``Integrity of {{Integrated Navigation Systems}},'' in \emph{2018 {{IEEE Conference}} on {{Communications}} and {{Network Security}} ({{CNS}})}, May 2018, pp. 1--5.

\bibitem{jmse11020267}
R.~Hopcraft, A.~V. Harish, K.~Tam, and K.~Jones, ``Raising the {{Standard}} of {{Maritime Voyage Data Recorder Security}},'' \emph{Journal of Marine Science and Engineering}, vol.~11, no.~2, p. 267, Jan. 2023.

\bibitem{Costin2023CybersecurityOC}
A.~Costin, S.~Khandker, H.~Turtiainen, and T.~H{\"a}m{\"a}l{\"a}inen, ``Cybersecurity of {{COSPAS-SARSAT}} and {{EPIRB}}: Threat and attacker models, exploits, future research,'' Feb. 2023.

\bibitem{osti_10166239}
F.~S. Roberts, D.~Egan, C.~Nelson, R.~Whytlaw, and {\relax CCICADA}.~Center, ``Combined {{Cyber}} and {{Physical Attacks}} on the {{Maritime Transportation System}},'' \emph{NMIOTC Maritime interdiction operations journal}, 2019.

\bibitem{9444397}
J.~Lee, Y.~Lee, D.~Lee, H.~Kwon, and D.~Shin, ``Classification of {{Attack Types}} and {{Analysis}} of {{Attack Methods}} for {{Profiling Phishing Mail Attack Groups}},'' \emph{IEEE Access}, vol.~9, pp. 80\,866--80\,872, 2021.

\bibitem{6497928}
M.~Khonji, Y.~Iraqi, and A.~Jones, ``Phishing {{Detection}}: {{A Literature Survey}},'' \emph{IEEE Communications Surveys \& Tutorials}, vol.~15, no.~4, pp. 2091--2121, 2013.

\bibitem{10.1007/978-981-19-6414-5_6}
G.~Potamos, S.~Theodoulou, E.~Stavrou, and S.~Stavrou, ``Building {{Maritime Cybersecurity Capacity Against Ransomware Attacks}},'' in \emph{Proceedings of the {{International Conference}} on {{Cybersecurity}}, {{Situational Awareness}} and {{Social Media}}}, C.~Onwubiko, P.~Rosati, A.~Rege, A.~Erola, X.~Bellekens, H.~Hindy, and M.~G. Jaatun, Eds.\hskip 1em plus 0.5em minus 0.4em\relax {Singapore}: {Springer Nature Singapore}, 2023, pp. 87--101.

\bibitem{9275344}
X.~Wang, J.~H. Park, H.~Liu, and X.~Zhang, ``Cooperative {{Output-Feedback Secure Control}} of {{Distributed Linear Cyber-Physical Systems Resist Intermittent DoS Attacks}},'' \emph{IEEE Transactions on Cybernetics}, vol.~51, no.~10, pp. 4924--4933, Oct. 2021.

\bibitem{8985406}
Y.~Ma, Z.~Nie, S.~Hu, Z.~Li, R.~Malekian, and M.~Sotelo, ``Fault {{Detection Filter}} and {{Controller Co-Design}} for {{Unmanned Surface Vehicles Under DoS Attacks}},'' \emph{IEEE Transactions on Intelligent Transportation Systems}, vol.~22, no.~3, pp. 1422--1434, Mar. 2021.

\bibitem{greenberg-2019}
\BIBentryALTinterwordspacing
A.~Greenberg, ``{The Highly Dangerous 'Triton' Hackers Have Probed the US Grid},'' 6 2019. [Online]. Available: \url{https://www.wired.com/story/triton-hackers-scan-us-power-grid/}
\BIBentrySTDinterwordspacing

\bibitem{6493150}
Y.~Yang, K.~McLaughlin, T.~Littler, S.~Sezer, E.~G. Im, Z.~Q. Yao, B.~Pranggono, and H.~F. Wang, ``Man-in-the-middle attack test-bed investigating cyber-security vulnerabilities in {{Smart Grid SCADA}} systems,'' in \emph{International {{Conference}} on {{Sustainable Power Generation}} and {{Supply}} ({{SUPERGEN}} 2012)}, Sep. 2012, pp. 1--8.

\bibitem{10069874}
K.~P. Swain, A.~Tiwari, A.~Sharma, S.~Chakrabarti, and A.~Karkare, ``Comprehensive {{Demonstration}} of {{Man-in-the-Middle Attack}} in {{PDC}} and {{PMU Network}},'' in \emph{2022 22nd {{National Power Systems Conference}} ({{NPSC}})}, Dec. 2022, pp. 213--217.

\bibitem{esrdc1314}
H.~Ravindra, J.~Langston, and K.~Schoder, ``{{MDD}} - {{System Model}} for {{RCPC Demonstration}} 1,'' \emph{ESRDC Website, www.esrdc.com}, 2022.

\bibitem{Riley2010}
\BIBentryALTinterwordspacing
G.~F. Riley and T.~R. Henderson, \emph{The ns-3 Network Simulator}.\hskip 1em plus 0.5em minus 0.4em\relax Berlin, Heidelberg: Springer Berlin Heidelberg, 2010, pp. 15--34. [Online]. Available: \url{https://doi.org/10.1007/978-3-642-12331-3_2}
\BIBentrySTDinterwordspacing

\bibitem{merkel2014docker}
D.~Merkel, ``Docker: lightweight linux containers for consistent development and deployment,'' \emph{Linux journal}, vol. 2014, no. 239, p.~2, 2014.

\bibitem{9917131}
T.-T. Nguyen, B.~L.-H. Nguyen, and T.~Vu, ``Resilience-{{Oriented Energy Management System}} for {{Ship Power Systems}},'' in \emph{2022 {{IEEE Power}} \& {{Energy Society General Meeting}} ({{PESGM}})}.\hskip 1em plus 0.5em minus 0.4em\relax {Denver, CO, USA}: {IEEE}, Jul. 2022, pp. 01--05.

\bibitem{9512317}
B.~L.~H. Nguyen, T.~Vu, C.~Ogilvie, H.~Ravindra, M.~Stanovich, K.~Schoder, M.~Steurer, C.~Konstantinou, H.~Ginn, and C.~Schegan, ``Advanced {{Load Shedding}} for {{Integrated Power}} and {{Energy Systems}},'' in \emph{2021 {{IEEE Electric Ship Technologies Symposium}} ({{ESTS}})}.\hskip 1em plus 0.5em minus 0.4em\relax {Arlington, VA, USA}: {IEEE}, Aug. 2021, pp. 1--6.

\bibitem{9512362}
M.~Sabah, I.~T. Ojo, and A.~M. Cramer, ``Evolution of {{Operability-Based Performance Metrics}} for {{Assessment}} of {{Mission Performance}},'' in \emph{2021 {{IEEE Electric Ship Technologies Symposium}} ({{ESTS}})}, Aug. 2021, pp. 1--6.

\bibitem{5705696}
A.~M. Cramer, S.~D. Sudhoff, and E.~L. Zivi, ``Performance {{Metrics}} for {{Electric Warship Integrated Engineering Plant Battle Damage Response}},'' \emph{IEEE Transactions on Aerospace and Electronic Systems}, vol.~47, no.~1, pp. 634--646, Jan. 2011.

\bibitem{Ledenev}
\BIBentryALTinterwordspacing
A.~Ledenev, ``Alexei-led/pumba: Chaos testing, network emulation, and stress testing tool for containers.'' [Online]. Available: \url{https://github.com/alexei-led/pumba}
\BIBentrySTDinterwordspacing

\bibitem{10.5555/1202316}
A.~Orebaugh, G.~Ramirez, J.~Beale, and J.~Wright, \emph{Wireshark \& Ethereal Network Protocol Analyzer Toolkit}.\hskip 1em plus 0.5em minus 0.4em\relax Syngress Publishing, 2007.

\bibitem{Sanfilippo_2022}
\BIBentryALTinterwordspacing
S.~Sanfilippo, ``Hping3: Kali linux tools,'' Aug 2022. [Online]. Available: \url{https://www.kali.org/tools/hping3/}
\BIBentrySTDinterwordspacing

\bibitem{8903954}
R.~R. S, R.~R, M.~Moharir, and S.~G, ``{{SCAPY- A}} powerful interactive packet manipulation program,'' in \emph{2018 {{International Conference}} on {{Networking}}, {{Embedded}} and {{Wireless Systems}} ({{ICNEWS}})}.\hskip 1em plus 0.5em minus 0.4em\relax {Bangalore, India}: {IEEE}, Dec. 2018, pp. 1--5.

\end{thebibliography}

\newpage




\vfill

\end{document}